\catcode`\@=11
\def\citen#1{\if@filesw \immediate\write \@auxout {\string\citation{#1}}\fi%
\@tempcntb\m@ne \let\@h@ld\relax \def\@citea{}%
\@for \@citeb:=#1\do {\@ifundefined {b@\@citeb}%
    {\@h@ld\@citea\@tempcntb\m@ne{\bf ?}%
    \@warning {Citation `\@citeb ' on page \thepage \space undefined}}%
    {\@tempcnta\@tempcntb \advance\@tempcnta\@ne
    \setbox\z@\hbox\bgroup\ifcat0\csname b@\@citeb \endcsname \relax
    \egroup \@tempcntb\number\csname b@\@citeb \endcsname \relax
    \else \egroup \@tempcntb\m@ne \fi \ifnum\@tempcnta=\@tempcntb
    \ifx\@h@ld\relax \edef \@h@ld{\@citea\csname b@\@citeb\endcsname}%
    \else \edef\@h@ld{\hbox{--}\penalty\@highpenalty
    \csname b@\@citeb\endcsname}\fi
    \else \@h@ld\@citea\csname b@\@citeb \endcsname \let\@h@ld\relax \fi}%
\def\@citea{,\penalty\@highpenalty\hskip.13em plus.13em minus.13em}}\@h@ld}
\def\@citex[#1]#2{\@cite{\citen{#2}}{#1}}%
\def\@cite#1#2{\leavevmode\unskip\ifnum\lastpenalty=\z@\penalty\@highpenalty\fi%
  \ [{\multiply\@highpenalty 3 #1%
  \if@tempswa,\penalty\@highpenalty\ #2\fi}]}   %
\makeatother 
\catcode`\@=12

\newcommand\Abs[1] {|#1\rangle\cy_{\!A}}
\def\ACY           {\mbox{${\mathfrak A}\cy$}}
\def\Ae            {\mbox{A$_\Id$}}
\newcommand\Aebs[1]{|#1\rangle\cy_{\!A_\Id}}

\def\Ainner        {\mbox{${\mathfrak A}_{\rm int}$}}
\def\alg           {algebra}
\def\Ao            {\mbox{A$_\omega$}}

\def\Atensor       {\mbox{${\mathfrak A}_{k_1k_2...k_r}$}}

\def\Awsusy        {\mbox{${\mathfrak A}_{\rm wsusy}$}}
\def\auto          {automorphism}

\def\bc            {boundary condition}
\def\Bc            {Boundary condition}
\def\be            {\begin{equation}}
\def\Be            {\mbox{B$^C$}}
\def\bearl         {\begin{array}{l}}
\def\bearll        {\begin{array}{ll}}
\def\bearlll       {\begin{array}{lll}}
\def\bfe           {{\bf1}}
\def\bs            {boundary state}

\def\cala          {{\mathfrak A}}

\def\calg          {{\cal G}}
\def\calgcy        {{\cal K}\cy}
\def\calgcyd       {K\cy}
\def\calgg         {{\cal K}\gep}
\def\calggd        {K\gep}
\def\calgp         {{\cal K}_{\omega_1\omega_2}}

\def\cals          {{\cal S}}
\def\calu          {{\cal U}}
\def\caly          {{\cal Y}}
\def\CCY           {\mbox{${\cal C}\cy$}}
\def\cft           {conformal field theory}

\def\cfts          {conformal field theories}
\def\Cgepner       {\mbox{${\cal C}\gep$}}
\def\chii          {\raisebox{.15em}{$\chi$}}
\def\Cinner        {\Cint}
\def\Cint          {\mbox{${\cal C}_{\rm int}$}}
\def\cla           {\mbox{$\mathfrak C$}}
\def\class         {classification}
\def\como          {coset model}
\def\compac        {compactification}
\def\complex       {{\dl C}}
\def\corfu         {correlation function}
\def\CP            {Chan\hy Paton }
\def\Cspacetime    {\mbox{${\cal C}_{\rm s{\sss-}t}$}}
\def\Ctensor       {\mbox{${\cal C}_{k_1k_2...k_r}$}}
\def\Cwsusy        {\mbox{${\cal C}_{\rm wsusy}$}}
\def\cy            {^{\sss\rm (CY)}}
\def\CY            {Cala\-bi\hy$\!$Yau }
\def\cym           {Cala\-bi\hy$\!$Yau manifold}

\def\dim           {dimension}
\def\dl            {\mathbb }
\def\dq            {\Delta q}
\def\ds            {\Delta s}

\def\dsty          {\displaystyle}
\def\ee            {\end{equation}}
\def\eE            {{\rm e}}
\def\eear          {\end{array}}
\def\eps           {\epsilon}
\def\eq            {\,{=}\,}
\newcommand\erf[1] {(\ref{#1})}
\newcommand\Erf[2] {(\ref{#1#2})}
\def\ext           {_{\rm ext}}
\def\extent        {\,\prec\,}
\def\findim        {finite-dimensional}
\newcommand\Frac[2]{\mbox{\large$\frac{#1}{#2}$}}
\def\furu          {fusion rule}
\def\futnote#1     {\footnote{~#1}\ }
\def\gep           {^{\sss(\rm Gep)}}
\def\gemo          {Gep\-ner model}
\newcommand\hsp[1] {\mbox{\hspace{#1 em}}}
\def\hy            {$\mbox{-\hspace{-.66 mm}-}$}
\def\id            {{\rm id}}
\def\Id            {{0}}
\def\ide           {identification}
\def\ii            {{\rm i}}
\def\iN            {\,{\in}\,}

\def\inner         {{\rm int}}
\def\irrep         {irreducible representation}

\newcommand\ishi[1]{|#1\rangle\!\rangle_{}}
\newcommand\ishio[1]{|#1\rangle\!\rangle_{\!\omega}^{}}
\def\J             {{\rm J}}
\def\L             {{\rm L}}
\long\def\labl#1   {\label{#1}\ee}
\long\def\Labl#1#2 {\label{#1#2}\ee}

\def\lie           {Lie algebra}
\def\Lie           {Lie group}
\def\llb           {\mbox{\large(}}
\def\Llb           {\mbox{\Large[}}
\def\LLb           {\mbox{\Large\{}}
\def\lrb           {\mbox{\large)}}
\def\Lrb           {\mbox{\Large]}}
\def\LRb           {\mbox{\Large\}}}
\def\Mapsto        {\,{\mapsto}\,}
\def\mimo          {minimal model}
\def\modinv        {modular invarian}
\def\Modinv        {Modular invarian}
\newcommand\N[3]   {{\rm N}_{#1,#2}^{\;\ \ #3}}
\newcommand\Ncy[3] {({\rm N}\cy)_{#1,#2}^{\ \ \ \ \ \ \ \ \ \ \ #3}}

\newcommand\Next[3]{({\rm N}\ext)_{#1,#2}^{\ \ \ \ \ \ #3}}

\newcommand\nxt[1] {\\\raisebox{.12em}{\rule{.35em}{.35em}}\hsp{.6}#1}
\def\nn            {$N\,{=}\,2$ }
\def\nnmimo        {$N\,{=}\,2$ minimal model}
\def\one           {\mbox{\small $1\!\!$}1}
\def\onedim        {one-dimen\-sional}

\def\ot            {\raisebox{.07em}{$\scriptstyle\otimes$}}
\def\oT            {\,\ot\,}
\def\otimeS        {\,{\otimes}\,}
\def\parfu         {partition function}
\def\psu           {{\hat\psi}}
\def\q             {quantum }
\def\Q             {Quantum }
\def\QB            {Q_{\scriptscriptstyle\rm BRST}}
\def\rcft          {rational conformal field theory}
\def\renog         {renormalization group}
\def\rep           {representation}
\def\resp          {respectively}
\def\rgs           {Ramond ground state}

\newcommand\sect[1]{\section{#1}\setcounter{equation}{0}}
\def\scs           {\scriptstyle}
\def\sce           {simple current extension}
\def\Scy           {(S\cy)}

\def\Sext          {(S\ext)}
\def\sss           {\scriptscriptstyle}
\def\exf           {{{\rm D}_{D/2+3}}}
\def\st            {\varpi}
\def\stot          {{\rm s}_{\rm tot}}
\def\stt           {string theory}

\def\suco          {superconformal }
\def\susic         {supersymmetric }
\def\susy          {supersymmetry}
\def\sutwo         {{\mathfrak{su}(2)}}
\def\sv            {{\rm u}}
\def\sw            {{\rm U}}
\def\syms          {sym\-me\-tries}
\def\tft           {topological field theory}
\def\Times         {\,{\times}\,}
\def\tS            {\tilde S}
\def\uone          {{\mathfrak u(1)}}
\def\vac           {\Omega}
\def\Vac           {\Phi^{0,0}_0}

\def\vv            {{\rm v}}
\def\wrt           {with respect to }
\def\wrtt          {with respect to the }
\def\ww            {{\rm w}}
\def\WZW           {Wess\hy Zu\-mino\hy Wit\-ten }
\def\wzwm          {WZW model}

\def\zet           {{\dl Z}}
\def\zetplus       {{\dl Z}_{>0}}

\documentclass[12pt]{article} \usepackage{amssymb,amsfonts,latexsym}

\begin{document}

\begin{flushright}  {~} \\[-1cm]
{\sf hep-th/0003298}\\{\sf PAR-LPTHE 00-09}\\{\sf ETH-TH/00-2}\\
{\sf CERN-TH/2000-0045}\\[1 mm]
{\sf March 2000} \end{flushright}

\begin{center} \vskip 14mm
{\Large\bf PROJECTIONS IN STRING THEORY AND}\\[4mm]
{\Large\bf BOUNDARY STATES FOR GEPNER MODELS}\\[20mm]
{\large J\"urgen Fuchs$\;^1$, \
Christoph Schweigert$\;^2$, \ Johannes Walcher$\;^{3,4}$}
\\[6mm]
$^1\;$ Institutionen f\"or fysik\\
Universitetsgatan 1\\ S\,--\,651\,88\, Karlstad\\[3.5mm]
$^2\;$ LPTHE, Universit\'e Paris VI\\
4 place Jussieu\\ F\,--\,75\,252\, Paris\, Cedex 05\\[3.5mm]
$^3\;$ Institut f\"ur Theoretische Physik \\
ETH H\"onggerberg\\ CH\,--\,8093\, Z\"urich\\[3.5mm]
$^4\;$ Theory Division, CERN\\CH\,--\,1211\, Gen\`eve 23
\end{center}
\vskip 18mm
\begin{quote}{\bf Abstract}\\[1mm]
In string theory various projections have to be imposed to ensure 
supersymmetry. We study the consequences of these projections in the 
presence of world sheet boundaries. A-type boundary conditions
come in several classes; only boundary fields that do not change the
class preserve supersymmetry. Our analysis takes in particular properly into 
account the resolution of fixed points under the projections. Thus e.g.\ the 
compositeness of some previously considered boundary states of Gepner models 
follows from chiral properties of the projections. Our arguments are model 
independent; in particular, integrality 
of all annulus coefficients is ensured by model independent arguments.
\end{quote}
\newpage


\sect{Introduction}

It has been appreciated for a long time that the construction of
consistent superstring theories requires appropriate projections
on the underlying \cft, most prominently the GSO projection. 
A careful implementation of these projections is e.g.\
required when one computes the massless spectrum of such theories.
A clear conceptual understanding becomes even more mandatory when
it comes to specifying boundary conditions for open strings.
The analysis of the interplay between projections in superstring theories 
and boundary conditions for open strings is our main concern in this paper. 
We will concentrate on compactifications of type II superstring theories
in which the internal part is an \nn rational \cft, among them in particular
the Gepner models.  Our approach enables us e.g.\ to derive formulas for
\gemo\ \bs s entirely from well-established principles. 
Where comparable with the literature, these results differ from 
the formulas obtained elsewhere except for some particularly simple models.

Let us be more explicit. In the construction of a superstring compactification
one starts with specifying the vacuum configuration. This amounts to choosing
a \cft\ \Cinner\ for the `inner' or `internal' sector. \Cinner\ must satisfy
a number of consistency constraints, such as possessing the correct Virasoro
central charge, enough supersymmetry on the world sheet, and modular invariance.
Afterwards, additional projections need to be imposed on \Cint. This includes 
in particular the GSO projection, which ensures space-time supersymmetry. But 
there is another generic projection, too, which in the case of flat backgrounds
looks quite innocent and which nevertheless will play an important role below.
Namely, the total \cft\ in question is a tensor product of the inner
sector with the flat space-time part, and the constraint that necessitates a
projection is that the spin structures for all fermionic fields on the world sheet 
must be aligned. In other words, the fermionic fields have to be either in the 
Ramond or in the Neveu\hy Schwarz sector simultaneously in each factor of the 
tensor product.  As we will see, the interplay between these two projections
has quite non-trivial consequences in non-flat backgrounds, in particular
for the description of boundary states.

Now it is well-known that just projecting out states from a \cft\ typically
destroys its consistency, like e.g.\ modular invariance of the torus
\parfu. The projection therefore must be compensated by some additional 
manipulations, such as including new, twisted, sectors. For instance, in 
the case of the Gepner \cite{gepn5} construction, the inner sector
\cft\ \Cinner\ one starts with can be written as a tensor product 
$\Ctensor\eq{\cal C}_{k_1}{\otimes}\cdots{\otimes}\,{\cal C}_{k_r}$
of \nn minimal models. On this theory \Ctensor\ one
imposes fermion alignment and the GSO projection, but at the same time
includes additional states that do not appear in the spectrum of the original
tensor product theory \cite{Gepn}. Put differently, the torus \parfu\ of the 
full Gepner \cft\ \Cgepner\ -- i.e.\ the theory that is obtained by these 
manipulations of projecting out old and of adding new states --
contains non-diagonal (`twisted') contributions when viewed in terms of 
primary fields of the \nn tensor product. In particular the vacuum field of the 
original theory \Ctensor\ not only gets combined with itself, but also with
other fields -- to be called {\em simple currents\/} -- of the original theory.
(The corresponding states play in fact a crucial role in the space-time
physics. The associated vertex operators provide e.g.\ the gravitini.)
At the chiral level, this means that the chiral symmetry
\alg\ of the Gepner model is {\em extended\/}
beyond that of the \nn tensor product, namely precisely by including the
relevant simple currents into the \alg\ \cite{scya4}. Note that from the 
point of view of the extended chiral \alg, the \parfu\ is diagonal.\,%
 \futnote{For simplicity, here we restrict ourselves to modular invariants of
 A-type for the \nn minimal models. Bulk spectra for other modular invariants
 have been computed in \cite{scya4,lysc3,fkss2}. For a recent discussion of
 boundary states in bulk theories with modular invariants of D-type or
 E-type (based on the results of \cite{bppz2}), see \cite{nano}.}

In short, the Gepner construction amounts to extending the underlying tensor 
product of \Cinner\ with the theory \Cspacetime\ that describes the surviving 
$D$ flat non-compact space-time dimensions by certain simple currents. 
For concreteness we refer to this new theory \Cgepner\ as the {\em Gepner
extension\/}. To separate generic aspects which are related to the
space-time part from aspects that depend on the chosen inner sector
it turns out to be simpler, and conceptually clearer, to break
up the extension into two separate steps. Thus we first perform a
suitable extension on the inner sector \Cinner\ alone. This way we arrive at
a theory to which we refer as the {\em \CY extension\/} \CCY. 
  In models that possess a geometric interpretation as a sigma model on a 
  \CY manifold, it is this theory
  \CCY, rather than the original theory \Cinner\ (e.g.\ the mere tensor product 
  \Ctensor\ of minimal models) that should be compared with the geometrical data. 
The proper combination of \CCY\ with the space-time theory \Cspacetime\
then still requires further projections. Thus in a second step, we tensor 
\CCY\ with \Cspacetime, and thereafter perform yet another extension that 
involves both \CCY\ and \Cspacetime. As we will see, this latter extension is 
completely straightforward. This allows us to concentrate our attention on \CCY.  

Simple current fields possess a variety of nice properties which allow for a 
very general and powerful treatment of arbitrary projections in which the 
chiral algebra gets enlarged \cite{scya,intr,scya6}. Such simple current 
extensions have often been compared to orbifold constructions.
For our purposes it is, however, indispensable not to mix up the
two operations of simple current extension and orbifolding. While the 
respective closed string partition functions indeed display a certain similarity --
both correspond to projecting out some states and adding new `twisted' states
-- there is a significant difference at the level of the chiral symmetry
algebras and, as a consequence, at the level of chiral \cft. Briefly,
in a \sce\ the chiral algebra $\cala$ gets {\em larger\/} \cite{scya6}
-- the new algebra $\cala\ext$ consists of the old one plus 
the simple current fields --  while in the orbifold construction it
gets {\em smaller\/} \cite{dvvv} -- the new algebra $\cala^G$ is the fixed point
{\em sub\/}algebra of the old one \wrtt orbifold group $G$.
Accordingly, a \sce\ of a given theory has, generically, fewer 
primary fields (inequivalent representations of the chiral
algebra) than the original theory; the `twisted states' that appear in 
the partition function correspond to left-right asymmetric combinations
of ordinary $\cala$-\rep s. On
the other hand, an orbifold has in general more primary fields than its
mother theory, and the additional states correspond to new fields which 
appear already at the chiral level
and carry `twisted representations' of the original chiral algebra $\cala$. 
The differences between the chiral aspects of the two constructions become
particularly relevant when it comes to the study of boundary effects. Still, 
these two types of constructing a new \cft\ from a given one are closely 
related -- they are in fact each other's inverse. The simple currents form 
an abelian group $\calg$ under the fusion product, and it can be shown
\cite{fuSc12} that the operations of extension by a group $\calg$ of 
simple currents and of taking the abelian orbifold \wrtt character
group $G\eq\calg^*_{}$ are precisely inverse to each other.\,%
 \futnote{It follows e.g.\ that the sizes of the stabilizer groups
 of the simple current and of the orbifold action are complementary,
 i.e.\ full simple current orbits correspond to orbifold fixed points
 and vice versa. For more quantitative statements, see section 7 of
 \cite{fuSc12}.}

The reason for emphasizing the differences between the various extended
theories that arise in a string compactification
rests in the following observation. Once one works with the
appropriate conformal field theory \CCY, the standard results for boundary
conditions in (unitary) \cfts\ can be employed. In particular, 
Cardy's \cite{card9} construction of \bs s for \bc s that preserve the full 
chiral \alg\ can be applied directly.

The main points of this paper are the following. After establishing the
necessary information about the Gepner and \CY extensions (section 2), in
section 3 we analyze in detail which symmetries of a Gepner model must be 
preserved and which ones can be broken by a given \bc. We also recall the recent 
increase of understanding of symmetry breaking \bc s (see \cite{fuSc10,fuSc11},
and also \cite{bifs} for applications to WZW models) and apply those results
to Gepner models. We thereby obtain all \bc s that preserve full \nn world 
sheet and half of space-time \susy, the so-called A-type \bc s. This includes
in particular the \bs s recently obtained in \cite{brSc}. Our analysis 
reveals that within the boundary conditions of A-type, different `automorphism
types' appear, so that the A-type conditions can be naturally partitioned 
into several subsets. Boundary operators that change the automorphism type 
of the boundary conditions do not respect the GSO projection and therefore, 
generically, describe unstable brane configurations. Explicit formulas for all 
\bs s of Gepner models which preserve the full extended algebra are given.

In section 4 we turn to boundary conditions for which the action of the 
chiral \alg\ of the inner sector \Cinner\ is twisted by some \auto, in 
particular the \Be-type conditions which are based on the mirror automorphism 
of the \nn \suco \alg. As the relevant chiral algebra of the 
Gepner model is larger than that of \Cinner, it is necessary to lift the 
automorphism to the simple current extension \CCY. 
In the analysis of both A- and B-type conditions we encounter the problem
that such a lift is typically not unique; we employ arguments from quantum 
Galois theory to describe this non-uniqueness (more details are provided
in the appendix). Finally, in section 5 we comment on the relationship 
between various ``singular'' structures encountered in Gepner models and 
their geometric counterparts and mention some open problems.

\sect{The Gepner extension and the \CY extension}

\subsection{The bosonic string map}

The simple current machinery was mostly developed for unitary
conformal field theories. But since for maintaining the world sheet \susy\
we must align the superghosts as well, the \cft\ of our interest is 
definitely not unitary. To deal with this problem, we make use of 
the {\em bosonic string map\/} \cite{enns,sche,gepn5,Lesw}. This stratagem
allows us to map the non-unitary chiral \cft\ of our primary interest to 
another chiral \cft\ that {\em is\/} unitary and that possesses the same 
topological data, i.e.\ modular matrices, but also braiding and 
fusing matrices. In particular, both in the open and closed string sector
we can then work with ordinary partition functions
rather than with supersymmetric partition functions. It is worthwhile to
point out that while the bosonic string map was originally designed to
construct heterotic theories, we use it here to simplify the description of
type II superstring theories which are supersymmetric both in the left and
the right chiral part.

Concretely, the fermions of the flat $D$-dimensional
space-time theory \Cspacetime\ together with the superghosts
can be described by the lorentzian lattice D$_{D/2,1}$; the first $D/2$
components come from the bosonization of the space-time fermions, and the
last one (with opposite sign in the kinetic energy) from the
bosonization of the superghost system. The bosonic string map $B$ then
amounts to replacing the non-unitary \cft\ D$_{D/2,1}$ 
by the unitary \cft\ D$_{D/2+3}$. Both of these theories have four primary
fields, corresponding to the four conjugacy classes of the D-type simple
\lie s; the map exchanges the characters for the zero ($o$) and vector ($v$)
conjugacy classes and multiplies the characters for the spinor ($s$) and 
conjugate spinor ($c$) conjugacy classes by $-1$. Thus $B$ is encoded in the 
matrix
  \be  B = \left(\begin{array}{rrrr}
  0&\one&0&0 \\ \one&0&0&0 \\ 0&0&-\one&0 \\ 0&0&0&-\one \end{array}\right) \,,
  \labl b
where $\one$ is a unit matrix in the state space of the additional
\cft\ with which the theory for fermions and superghosts gets tensored, i.e.\
of the inner sector theory  \Cinner\ and the bosonic part of \Cspacetime.
Denoting by a tilde quantities before the string map
(`supersymmetric quantities'), and without tilde the ones after the string map
(`ordinary CFT quantities'), we thus have, schematically, $\tilde{\chii}\eq
B\chii$. For the modular transformation matrices this amounts to
  \be  \tilde S = B\, S\, B^{-1}\,,  \qquad \tilde T = B\, T\, B^{-1} \,. 
  \Labl st
Given the modular invariant torus \parfu\ $Z$ of the ordinary \cft, which
satisfies $[Z,S]\eq0\eq[Z,T]$, it follows immediately that
  \be  \tilde Z := B\, Z\, B^{-1} \ee
is modular invariant on the supersymmetric side.

The \nn superconformal algebra contains a $\uone$ current sub\alg.
Via spectral flow, space-time \susy\ is achieved when all $\uone$ charges 
\wrt this sub\alg\ are odd integers. This condition can be fulfilled by a
suitable projection on the allowed \rep s; this is precisely what the
GSO projection does. Because of the exchange between $o$ and $v$ and the 
$r$-dependence of the $\uone$ charge of $D_r$-spinors, the bosonic string
map \erf b changes all charges \wrtt $\uone$ sub\alg\ of the \nn \alg\
by $1\bmod2$ \cite{Gepn}. This means in particular that while in the 
supersymmetric theory the GSO projection is to {\em odd\/} integral $\uone$
charges, in the bosonic theory it is to {\em even\/} integral $\uone$ charges.

\subsection{Simple current extensions}

Starting from the tensor product theory $\Cspacetime\otimeS\Cinner$,
the Gepner construction proceeds by projecting out certain states and adding 
new ones \cite{Gepn}.  As already mentioned, technically this can be realized 
by the procedure of {\em simple current extension\/}. Basically, the \sce\ of
a \cft\ with chiral algebra $\cala$ by some group $\calg$ of simple currents 
of integral conformal weight has the following effects \cite{scya,scya6,fusS6}.
\nxt
When fused with any other primary field $\lambda$ of the theory, a simple 
current $\J$ yields just a single field $\J\lambda$.
Thus a simple current is invertible in the fusion ring.
The group $\calg$ then acts on the fusion ring of the $\cala$-theory, and
the \sce\ amounts to dividing out this action of $\calg$.
\nxt
The {\em projection\/} amounts to keep only those fields $\lambda$ which obey
$Q_\J(\lambda)\eq0$ for all $\J\iN\calg$, where
  \be  Q_\J(\lambda) := \Delta_{\lambda}+\Delta_\J -\Delta_{\J\lambda}\bmod\zet
  = \Delta_{\lambda}-\Delta_{\J\lambda}\bmod\zet  \Labl mc
is the so-called monodromy charge of the field $\lambda$ of
the $\cala$-theory with respect to the simple current (with integral
conformal weight) $\J\iN\calg$.
\nxt
To obtain the primary fields of the extended theory we must organize the
$\cala$-fields that survive the projection into orbits $[\lambda]$ under the
fusion product with the currents in $\calg$.
\nxt
The diagonal modular invariant of the extended theory reads
  \be  Z\ext =
  \sum_{[\lambda] \atop Q_\J(\lambda)=0 \;\forall \J\in\calg}\!\!
  |\cals_\lambda|\, {\mbox{\Large$|$}} \!\!\!\sum_{\J\in\calg/\cals_\lambda}
  \!\chii^{}_{\J\lambda}(\tau){\mbox{\Large$|$}}^2 \,,  \Labl7e
where $\cals_\lambda\,{\subseteq}\,\calg$ is the so-called
stabilizer of $\lambda$, i.e.\ the subgroup $\cals_\lambda$ consisting of those
elements of $\calg$ which leave $\lambda$ fixed under the fusion product of the
$\cala$-theory. Note that \Erf7e is {\em non-diagonal\/} when viewed in terms of
the primaries of the original theory. The terms of the form 
$\chii_\lambda^{} \chii_{\J\lambda}^*$ indicate the inclusion of 
{\em twisted\/} states which are needed to ensure modular invariance.
(For an analysis in the WZW case, see \cite{fegk}.) Also, both the stabilizer 
subgroup $\cals_\lambda$ and the monodromy charge are well defined for orbits
$[\lambda]$, not only for individual fields $\lambda$.
\nxt
When an orbit $[\lambda]$ has a non-trivial stabilizer, the factor of
$|\cals_\lambda|$ in the partition function \Erf7e seems to indicate that
the corresponding states occur several times. An additional `quantum number'
distinguishing those states is provided by a character of
$\cals_\lambda$, i.e.\ by $\psi_\lambda\iN\cals_{\lambda}^*$. Accordingly, the
primary fields of the extended theory are completely labeled as
  \be  \lambda\ext = [\lambda,\psi_\lambda] \,.  \Labl lx

It is worth stressing that, while the prescription for projecting out states 
and adding new ones is already in itself sufficient for obtaining the spectrum 
of the model, to determine the complete modular properties of the model (and 
a fortiori for obtaining boundary conditions) it is indispensable to 
take proper care of such additional quantum numbers. Naively,
in the case of a non-trivial stabilizer the projection rules appear to require
the inclusion of the same state several times into the partition function.
This would spoil unitarity of the modular S-matrix of 
the theory. The puzzle is resolved by realizing that those seemingly
identical states are indeed distinguished by a further quantum number. 
Simple currents constitute a convenient conceptual framework for
summarizing the required additional information.
\nxt
The modular S-matrix $S\ext$ of the extended theory can be expressed 
in terms of the modular S-matrix $S$ of the $\cala$-theory and of 
similar matrices $S^\J$ with $\J\iN\calg$. The latter describe the modular 
S-transformation of one-point chiral blocks (of the $\cala$-theory) on the 
torus with insertion $\J$ \cite{fusS6,bant7}. Explicitly \cite{fusS6,fuSc10},
the matrix elements of $S\ext$ (labeled, according to the above, by
$\calg$-orbits of monodromy charge zero $\cala$-primaries $\mu$,
supplemented by a character $\psi_\mu$ of the stabilizer $\cals_\mu$) read
  \be  \Sext_{[\lambda,\psi_\lambda],[\mu,\psi_\mu]}^{}
  = \Frac{|\calg|}{|\cals_\lambda|\,|\cals_\mu|}
  \sum_{\J\in\cals_\lambda\cap\cals_\mu} \psi_\lambda(\J)\,
  \psi_\mu(\J)^*\, S^\J_{\lambda,\mu} \,,  \Labl xS
where $S^\vac\,{\equiv}\,S$ is the ordinary S-matrix.
When there are no fixed points (i.e., orbits 
with non-trivial stabilizer), then this expression collapses to 
  \be  \Sext_{[\lambda],[\mu]}^{} = |\calg|\, S_{\lambda,\mu} \,,  \Labl0f
so in this particular case the original S-matrix already contains all 
information about $S\ext$.

Actually the formula \Erf xS does not cover the most general situation. 
In full generality we rather have to account for the fact that the 
implementation of symmetries in quantum systems is typically only projective.
This can also happen for the symmetries studied here.  Quantitatively, the 
effect is described by a two-cocycle on the stabilizer group $\cals_\mu$. What 
is remarkable is that this two-cocycle can be computed entirely in terms of the
matrices $S^\J$. One can show \cite{fusS6} that the projectivity 
is properly taken into account by replacing $\cals_\mu$ by the subgroup
  \be  \calu_\mu\,\subseteq\,\cals_\mu  \Labl us
of $\cals_\mu$ on which the two-cocycle vanishes; $\calu_\mu$
is called the untwisted stabilizer of $\mu$.\,%
 \futnote{The formula for the extended S-matrix then reads
  $$ \Sext_{[\lambda,\psu_\lambda],[\mu,\psu_\mu]} = |\calg|\,
  [\, |\cals_\lambda|\,|\calu_\lambda|\,|\cals_\mu|\,|\calu_\mu| \,]_{}^{-1/2}
  \!\sum_{\J\in\calu_\lambda\cap\calu_\mu}\! \psu_\lambda(\J)\,
  \psu_\mu(\J)^*\, S^\J_{\lambda,\mu} \,, $$
 where $\psu_\mu$ is a character
 of $\calu_\mu\,{\subseteq}\,\cals_\mu\,{\subseteq}\,\calg$.
 Note that in this general situation even the labeling of primary fields
 is different from the case where $\calu_\mu$ coincides with $\cals_\mu$
 for all $\mu$; in place of the label \Erf lx we now have
 $\lambda\ext\eq[\lambda,\psu_\lambda]$.
 Our notation for the simple current orbits is actually adapted to the
 general situation, as $\calg_\mu/\calu_\mu$ acts non-trivially on the
 characters $\psu_\mu$ when $\calu_\mu$ is a proper subgroup of $\cals_\mu$.
\\
 Also, while the arguments in \cite{fusS6} were not sufficient to prove
 the formula rigorously, a proof is possible by combining them with the
 results of \cite{brug2} on the uniqueness of the modularisation of a 
 premodular category. Independently, various aspects of the formula
 can be tested directly \cite{fusS6,bant7}. For instance, with the help of the 
 computer program {\tt kac} (see {\tt http:/$\!$/norma.nikhef.nl/%
 \raisebox{-.2em}{$\tilde{\phantom.}\,$}t58/kac.html})
 it was checked in a huge number of cases that
 it produces non-negative integers when inserted into the Verlinde formula.
 Moreover, manifestly the formula requires only information about the chiral
 \cft, and even only information about topological aspects of the chiral
 theory. In particular it does not involve any knowledge about \bc s.
\\
 While in \gemo s with diagonal or charge conjugation invariant one
 always has $\calu_\mu\eq\cals_\mu$, for models
 where the extension of the \nn tensor product is by a larger group -- e.g.\
 corresponding to taking non-diagonal modular invariants of the \nn
 minimal models -- cases where $\calu_\mu$ is a proper subgroup of
 $\cals_\mu$ can and do arise. This must e.g.\ be taken into account 
 when analyzing the boundary states introduced in \cite{nano}.\label{futn}}
In \gemo s with diagonal (or charge conjugation) torus \parfu\ -- the situation
of our main interest below -- one always has $\calu_\mu\eq\cals_\mu$, so that
henceforth we will ignore this modification.

Via the Verlinde formula, the fusion rules of the extended theory can then
be expressed through the fusion rules of the $\cala$-theory and the fixed 
point quantities $\cals_\lambda$ and $S^\J$.
For instance, for $\calg\eq\zet_2\eq\{\vac,\J\}$ one finds
  \be \bearll
  \Next{[\lambda,\psi]}{[\lambda',\psi']}{[\lambda'',\psi'']} \!\!
  &= \Frac1{|\cals_\lambda^{}|\,|\cals_{\lambda'}|\,|\cals_{\lambda''}|}\,
  \Llb\, \N\lambda{\lambda'}{\lambda''} + \N\lambda{\J\lambda'}{\;\lambda''}
  + 2 \dsty\sum_{\mu\atop \J\mu=\mu}
  \llb \psi'\psi''\, \Frac{S^{}_{\lambda,\mu}S^\J_{\lambda',\mu}
                     S^{\J*}_{\lambda'',\mu}} {S_{\vac,\mu}}
  \\{}\\[-.9em]&\hsp{9.3}
     + \psi\psi''\,  \Frac{S^\J_{\lambda,\mu}S^{}_{\lambda',\mu}
                     S^{\J*}_{\lambda'',\mu}} {S_{\vac,\mu}}
     + \psi\psi'\,   \Frac{S^\J_{\lambda,\mu}S^\J_{\lambda',\mu}
                     S^{{}*}_{\lambda'',\mu}} {S_{\vac,\mu}}  \lrb \Lrb
  \,. \eear \Labl{z2fp}
As is clear from \Erf xS, the rows and columns of the $S^\J$-matrices are 
labeled by only those primaries of the $\cala$-theory that are fixed under 
$\J$. Thus unless at least two of the fields $\lambda$, $\lambda'$ and
$\lambda''$ are fixed under $\J$, the corresponding terms in \erf{z2fp} vanish.

It is worth emphasizing that a simple current extension amounts to
nothing else than to a change of the underlying {\em chiral\/} \cft. This 
fact, which is somewhat hidden in other treatments of projections (compare
e.g.\ \cite{vafa4}), is of central importance for 
gaining a better understanding of boundary states in 
Gepner models. Namely, it implies in particular that all the features that are
revealed in the analysis of these projections can be understood
in a manner that is completely independent from our (yet uncomplete)
understanding of world sheet boundary effects. Since the relevant results 
have undergone extensive physical and mathematical consistency checks which do 
not use any information about boundary conditions, we can safely exploit these 
structures as an input in the construction of boundary states for Gepner models.

A crucial property of superconformal field theories is that the supersymmetry 
automatically leads to the existence of certain simple currents. First of all, 
independently on the number of world sheet supersymmetries, every \suco field 
theory has a distinguished simple current $v$: the generator of world sheet 
\susy, which has order two and conformal dimension $\Delta\eq 3/2$. The 
monodromy charge \wrt $v$ is 0 for primary fields in the Neveu\hy Schwarz 
sector and 1/2 for primaries in the Ramond sector. The `superpartner' of a 
primary field $\lambda$ is given by fusion product of $\lambda$ with this 
simple current $v$.

In case the superconformal field theory has extended ($N\eq2$) supersymmetry,
there is yet another simple current $s_\inner$: the \rgs\ $R_0$ with highest 
$\uone$ charge. This can be seen after expressing the $\uone$ current $J$ of 
the \nn algebra in terms of a canonical free boson, $J(z)\eq\ii\sqrt{c/3}\,
\partial\phi(z)$. Then the \rgs\ is given by $\exp(\ii \sqrt{c/12}\,\phi)$. 
Its conformal dimension is $\Delta\eq c/24$, as befits a \rgs, and it has the
correct $\uone$ charge $c/6$. In this formulation, it is easy to see that the
monodromy charge \wrt this simple current equals half of the superconformal
$\uone$ charge of a field. 

\subsection{The Gepner extension}

In this subsection we display the simple current extension that leads from 
an internal \nn superconformal theory \Cinner\ to a consistent string background
\Cgepner. Let us point out that this extension is not only applicable for the 
original Gepner models, but likewise for any other \nn compactification in which 
the internal theory is a rational \cft, for instance for Kazama\hy Suzuki models.

The (flat) space-time bosons and Virasoro ghosts will play no role in what
follows, and accordingly we suppress their contribution to \Cspacetime.
What then remains of the space-time theory are the fermions and superghosts.
After the bosonic string map, these are described by 
the level one WZW theory based on the Lie algebra D$_{D/2+3}$; 
this (unitary) \cft\ $\exf$ has four primary fields, which we label as
   \be  \st \in \{ o,s,v,c \} \,.  \labl{osvc}
The Gepner extension is the extension of the tensor product theory 
$\exf\otimeS\Cinner$ by a certain simple current group $\calgg$, which
accounts for fermion alignment and GSO projection. The group $\calgg$
is generated by some number $r$ of order-two currents
  \be  \vv_i := (v;v_{i,\inner})  \Labl0v
together with 
  \be  \stot:= (s;s_\inner) \,;   \Labl0s
the fields $\vv_i$ will be referred to as {\em alignment currents\/}
(rather than as vector currents, as is done e.g.\ in \cite{sche3}), and $\stot$
as the {\em total spinor current\/}. When the inner 
sector \Cinner\ can itself be written as a tensor product (e.g.\
of \nn \mimo s as in the original Gepner construction), then each tensor
factor provides us with one of the currents $v_{i,\inner}$, which then is a
non-trivial field in the $i$th factor, tensored with the identity field
of all other factors of \Cinner. The subspace with lowest conformal weight of 
the \rep\ space of the bosonic part of the \nn \alg\ corresponding to the 
field $v_{i,\inner}$ has a dimension which is a multiple of two; it 
contains the two supercurrents $G_i^\pm$ of the $i$th tensor factor.

The order $N_s$ of $\stot$ is model dependent. Also, depending on the model 
the resulting group $\calgg$ is either the direct product of the $\zet_{N_s}$ 
generated by the total spinor current $\stot$ and of the $\zet_2$ groups 
generated by the alignment currents $\vv_i$, or else some quotient of that 
direct product group. The latter happens when the field $(\stot)^{N_s/2}$
is contained in $(\zet_2)^r$,\,%
 \futnote{For instance, in \nn minimal models with odd level $k$, the spinor
 current $s$ (see formula \Erf3s below) satisfies $s^{2k+4}\eq v$.
 This results in the equality $\stot^{N_s/2}\eq\prod_{i=1}^r\vv_i$ when
 \Cinner\ is the tensor product of $r$ \mimo s with only odd levels.
 For more details, see subsection \ref{ssmm}.}
and in that case the corresponding quotient is, as an abstract abelian group,
the direct product of $\zet_{N_s}$ with $r{-}1$ copies of $\zet_2$. Thus 
the simple current group of the Gepner extension has the structure
  \be  \calgg = \zet_{N_s}^{} \times (\zet_2)^{r-\eta}_{} \quad{\rm with}\quad
  \eta\iN\{0,1\}\,.  \Labl gg
The extension by the group generated by the currents $\vv_i$ guarantees world 
sheet supersymmetry. Indeed, the total \nn superconformal algebra must split
into two modules of its bosonic subalgebra, the vacuum module and
a module containing the supercurrents $G^\pm\eq\sum_{i=1}^r G_i^\pm$; this 
is so only {\em after\/} extension by the $\vv_i$. (Concretely, e.g.\ the two 
terms in $G_1^\pm{+}G_2^\pm\,{\equiv}\,G_1^\pm\oT\bfe\,{+}\,\bfe\oT G_2^\pm$
lie in two distinct irreducible modules of \Ainner, and these modules get
combined into an irreducible module of the extended \alg\ precisely due to
the extension by $\vv_1\vv_2\eq (o;v_{1,\inner}v_{2,\inner})$.)
The extension by the total spinor current $\stot$ implements 
the GSO projection and hence ensures space-time supersymmetry. The monodromy 
charge \Erf mc \wrtt current $\stot$ can be shown to coincide (modulo $\zet$) 
with half of the superconformal $\uone$ charge of a state. Also, a change 
from the $o$ to the $v$ conjugacy class results in a change of this monodromy 
charge by $1/2\bmod\zet$, and in the Ramond sector the same effect results 
from the $r$-dependence of the $\uone$ charge of the spinors of $D_r$. 
Recalling the form \erf b of the bosonic string map, we thus see again that 
it changes the effect of the GSO projection from a projection to odd
integral $\uone$ charges to a projection to even integral $\uone$ charges.

Let us remark that the abelian orbifold construction that brings us
back from the \gemo\ \Cgepner\ to the tensor product $\exf\otimeS\Cinner$
consists of orbifolding by the group $\calggd$ that is generated by the \auto s
  \be  \bearl \left. \bearl
  \vv_{i}\Mapsto{-}\vv_{i}\,, \ \
  \vv_j\Mapsto \vv_j\;\mbox{ for }j\,{\ne}\,i \\
  \stot\Mapsto \stot\,, \eear\right\} \mbox{  for }i\eq1,2,...\,,r\,,
  \qquad{\rm and} \\{}\\[-.6em] \hsp{.5}
  \vv_j\Mapsto \vv_j\;\mbox{ for all }j\,,\quad
  \stot\Mapsto\exp(2\pi\ii/N_s)\,\stot \,. \eear   \labl K
Also note that the latter map corresponds to a shift $\phi\Mapsto\phi{+}
4\pi\sqrt{3/c}/N_s$ of the free boson $\phi$ in terms of which the spinor 
current can be written as $\stot\eq\exp(\ii \sqrt{c/12}\,\phi)$.

\subsection{The \CY extension}

The original Gepner construction -- reformulated in the previous subsection 
in terms of simple current extensions -- involves both the flat space-time part 
$\exf$ and the inner sector \Cinner. This 
tends to obscure the connection with the geometric formulation in terms 
of sigma models on \CY manifolds. The object in the Gepner model that 
corresponds to the compactification manifold in the geometric setting is not 
simply the \cft\ \Cinner\ -- e.g.\ the tensor product \Ctensor\ of \mimo s --
but rather an extension of this tensor product, which will be specified
shortly; we shall denote it by \CCY\ and call it the {\em \CY extension\/}.
The connection to geometric compactifications is usually derived using
a Landau\hy Ginzburg description of the \mimo s
(for a different line of arguments, see \cite{nawe}), and indeed this
construction involves a non-trivial projection on the tensor product,
commonly referred to as forming a Landau\hy Ginzburg orbifold \cite{vafa4}.
In a second step, one combines this extended inner sector \CCY\ with flat 
space-time (i.e.\ tensors with the theory $\exf$) and then performs an
additional extension, which has more similarities to the GSO projection in 
ten flat dimensions. Unlike the step from \Cinner\ to \CCY, this further 
extension is completely straightforward.  Let us stress that the procedure 
that we call the \CY extension can be performed independently of any 
connection of the internal sector to a classical geometrical compactification.

Inspecting the Gepner extension, one observes that the group $\calgg$ 
contains many currents that have a trivial space-time part and therefore 
effectively define an extension of the internal theory \Cinner\ alone. 
We denote the group of these currents by $\calgcy$. 
Using the fusion rules of the $\exf$ theory (which are of 
the form $\zet_2\Times\zet_2$ when we compactify to $D\eq d+2\eq6$ dimensions,
and $\zet_4$ for compactifications to $D\eq4$ or $8$), we find that $\calgcy$ 
is generated by all products of any two of the currents $v_{i,\inner}$, and hence by
  \be  \ww_i := v_{1,\inner}\,v_{i,\inner}\,, \qquad i\iN\{2,3,...\,,r\} \ee
which will again be called alignment currents, together with the current\,%
 \futnote{The presence of the power of $v_{1,\inner}$ accounts
 for the fact that $s^2\eq v^{d/2}$ in the $\exf$ theory. More explicitly,
 for compactifications to $D\eq6$, the group $\calgcy$ contains all
 products of an even number of currents $v_{i,\inner}$ and all products of
 $(s_\inner)^2$ with the former, while for compactifications to $D\eq4$ or 8
 in addition to all products of an even number of currents $v_{i,\inner}$
 one has all products of $(s_\inner)^2$ with an odd number of $v_{i,\inner}$.} 
  \be  \sv := s_\inner^2 (v_{1,\inner})_{}^{d/2} \,.  \ee
This group has the structure
  \be  \calgcy = \zet_{N_s/2}^{} \times (\zet_2)^{r-1-\eta}_{} \,, \labl{gcy}
where the contribution $\eta\iN\{0,1\}$ in the exponent accounts for the
possibility that $\sv^{N_s/4}_{}$ is 
contained in the product of the $r{-}1$ $\zet_2$ groups that are generated by
the currents $\ww_i$ (compare the remarks before formula \Erf gg).

The theory \CCY\ that is obtained upon extension of \Cinner\ by $\calgcy$ inherits
a simple current (sub)group $\{o\cy,s\cy,v\cy,c\cy\}$ that has the same fusion 
rules (i.e., $\zet_2\Times\zet_2$ and $\zet_4$, \resp) as the $\exf$ theory. 
The final projections that bring us from $\exf\otimeS\CCY$ to the Gepner extension 
\Cgepner\ amount to extending by the simple current $(v;v\cy)$, which aligns 
fermions in $\exf$ and \CCY, and in addition by either $(s;s\cy)$ or $(c;s\cy)$.
This very last extension is the true analog of the GSO projection in flat 
space-time. In particular, when one combines the left and right halves of 
the theory, the choice between type IIA and IIB theories is 
equivalent to a choice between the extension by $(s;s\cy)$ both on the left
and the right (or equivalently, by $(c;s\cy)$ on both sides), or else
by $(s;s\cy)$ on one side and by $(c;s\cy)$ on the other. We remark that
all currents in the extension of $\exf\otimeS\CCY$ to \Cgepner\ act freely.
Thus the simple formula \Erf0f for the modular S-matrix of the extension applies,
and hence as announced this extension is straightforward.

The dependence of the precise structure of the
$\calgcy$ extension on the number of compactified dimensions
can be traced back to the fact that the internal spectral flow operator, 
mapping the R to the NS sector, changes the $\uone$ charge by $c_\inner/6$, 
where $c_\inner\eq 12-3d/2$ is the central charge of $\Cinner$. 
Thus, while in the NS sector we always project onto integral internal $\uone$ 
charge, the internal $\uone$ charges in the R sector are either integers 
or in $\zet{+}1/2$, depending on the external dimension
being $6$ or $4,8$, respectively. The integrality of the $\uone$ charges in 
the NS sector is necessary to make a correspondence between chiral primary 
fields and differential forms in the geometric compactification possible,
and is therefore highly welcome. In fact, $s_\inner^2$ can be identified 
model independently as the square of the \rgs\ with maximal $\uone$ charge.

As we shall see below, there is yet another intermediate theory, to be denoted 
by \Cwsusy, that is of interest. This is the theory 
that one obtains from the inner sector \Cinner\ by extending with the 
subgroup $(\zet_2)^{r-1}{\subset}\,\calgcy$ generated by the alignment
currents $\ww_i$ only, i.e.\ by enforcing only
fermion alignment, and hence world sheet \susy, on the internal theory. 
\Cwsusy\ will play an important role in the analysis of boundary conditions.
We summarize the relation between the various extensions schematically as
  \be
  \exf\otimeS\Cinner\extent\exf\otimeS\Cwsusy\extent\exf\otimeS\CCY\extent
  \Cgepner \,.  \labl{extent}
Note that the group that furnishes the extension from \Cwsusy\ to \CCY\
is the cyclic group generated by the image $\sw$ of the simple current
$\sv$ \Erf sv in the \Cwsusy-theory. The order of $\sw$ can differ from 
the order $N_s/2$ of $\sv$ by a factor of two; it is given by $N_s'/2$ with
  \be  N_s' = 2^{-\eta}_{} N_s \,,  \Labl uu
where $\eta$ is the integer introduced in formula \Erf gg.

We can -- and will -- simplify the discussion and restrict our attention in
the sequel to the intermediate theory \CCY\ (and later on also to \Cwsusy). 
As the additional extension to \Cgepner\ is so simple, we do not loose 
any essential features when doing so. In particular the issue of fixed 
points arises always only in the study of \CCY. To conclude this section, let us
emphasize that the construction described above is model independent and
does not rely on specific aspects of the \nn super\cft\ used in the inner sector.

\sect{A-type boundary conditions}

\subsection{Intermediate chiral algebras}\label{sica}

Already in closed string theory the simple current extension leading to 
\CCY\ must be taken into account properly. In particular, a careful 
treatment of fixed points is compulsory to find the correct massless spectrum, 
compare e.g.\ \cite{sche3,fuSc}. When open strings are present, for the 
following reason an even deeper understanding of fixed point resolution is 
required. In the computation of the massless spectrum one just
counts states, thus a detailed understanding of the underlying partition
functions suffices. In contrast, as was first realized by Cardy \cite{card9},
in the construction of boundary states the modular S-matrix enters directly;
therefore a complete knowledge of this matrix in simple current extensions 
is required as well.
Moreover, open string partition functions (annulus amplitudes) also implicitly
contain the modular matrices, so that not even the open string {\em spectrum\/}
can be obtained correctly without proper resolution of the fixed points.
In this context, it is an important observation that typically 
the \CY extension, and hence also the Gepner extension,
do possess fixed points. For instance, as will be discussed in subsection 
\ref{ssmm}, in the case of a tensor product $\Cinner\eq\Ctensor$ of
\nnmimo s, fixed points arise precisely if at least one level $k_i$ is even. 

The group of automorphisms of the \nn superconformal algebra is the Lie group
O$(2)$. This group has two connected components, and any element of the
component not connected to the identity can be obtained by composing an
element of the identity component with the mirror \auto\ (see formula \Erf41).  
Those automorphisms of the chiral \alg\ of a tensor product of internal models 
that respect the \nn structure are generically given by O$(2)$ as well (but 
additional permutation symmetries are present when some of the factors of the 
tensor product are identical).  Accordingly, in such string compactifications
one conventionally distinguishes between two classes of \bs s:
Those which correspond to an \auto\ in the identity component of O$(2)$,
and those corresponding to an \auto\ in the other component. In the literature, 
the former states are often collectively referred to as 
A-{\em type \bs s\/}, while latter are said to be of B-{\em type\/}.
A-type \bc s leave the chiral \alg\ \Ainner\ of the inner sector \Cinner\
invariant; insisting that also an \nn subalgebra of the extension \CCY\
remains unbroken, the A-type automorphism must be the identity map.  
In this section we study in detail \bc s $\Abs a$ which do preserve \Ainner,
i.e.\ which satisfy
  \be  \llb Y_n\oT\bfe + (-1)^{\Delta_Y-1}\, \bfe\oT Y_{-n} \lrb
  \Abs a = 0  \Labl8y
for every field $Y(z)\eq\sum_{n\in\zet}Y_n z^{-n-\Delta_Y}$ of
conformal weight $\Delta_Y$ in \Ainner.

We start our analysis
by recalling that the total chiral \alg\ \ACY\ of the \CY extension \CCY\
is obtained from \Ainner\ by a simple current extension with the group
$\calgcy$. A generic \bs\ $\Abs a$ will not preserve all of \ACY, but 
only some sub\alg\ $\cala_a$ containing \Ainner. This sub\alg\ cannot be 
arbitrary, though. First of all, we are interested in conformally invariant 
boundary conditions only, and hence the Virasoro subalgebra of \ACY\ must be 
preserved.  This is automatically satisfied by the \bs s $\Abs a$, since the
Virasoro algebra is already contained in the inner sector algebra \Ainner.
But in addition the preserved subalgebra must have enough structure to allow 
for the construction of conformal blocks and, based on them, of correlation 
functions. One therefore has to require that $\cala_a$ must again be a
vertex operator \alg. To get an overview over all 
boundary conditions of our present interest, we thus look for all 
vertex operator algebras $\cala$ that lie between $\Ainner$ and $\ACY$:
  \be  \Ainner \,\subseteq\, \cala \,\subseteq\, \ACY \, . \ee

For general vertex operator algebras, this \class\ of sub\alg s would be a 
hopeless problem.  Here, however, we know that the inner sector chiral \alg\ 
\Ainner\ can be characterized as the sub\alg\ of \ACY\ that is left pointwise 
fixed by the group $\calgcyd$ of automorphisms of $\ACY$, described in 
formula \erf K, which is dual to $\calgcy$. This observation allows us
to employ a basic result from the Galois theory for vertex operator
algebras \cite{doMa2,doMa5,hamt}, which tells us that the possible
chiral algebras between \Ainner\ and \ACY\ are in one-to-one
correspondence with subgroups of the group $\calgcyd$. 
We conclude that to every \bs\ $\Abs a$ of \CCY\ we can associate a 
subgroup $\calgcyd_a$ of $\calgcyd$, such that the sub\alg\ of \ACY\ that is
preserved by the \bc\ is the fixed point \alg\ \wrt $\calgcyd_a$.

\subsection{Boundary states and \auto\ types}

Boundary conditions that preserve a fixed point \alg\ of the bulk symmetries
\wrt some finite abelian group of \auto s have been studied in 
\cite{fuSc11,fuSc12}. When applied to the present situation, the pertinent 
results of \cite{fuSc11,fuSc12} may be summarized as follows.
\nxt
Using notation from the unextended theory \Cinner, the \bs s can be 
labeled in a way much similar to the labeling \Erf lx of primary fields of
the extended theory \CCY, namely as
  \be  a = [\mu,\psi_\mu] \,.  \ee
The difference is that, unlike in \Erf lx, here $\mu$ can be any
primary field label of \Cinner, i.e.\ now there is no restriction on the
monodromy charge.\,%
 \futnote{At this point, the possibility of having untwisted stabilizers 
 $\calu_\mu\,{\subset}\,\cals_\mu$ (see formula \Erf us) must in general 
 be taken into account.
 Then $\psi_\mu$ gets in fact replaced by a character $\psu_\mu$ of the 
 untwisted stabilizer $\calu_\mu$, and the simple current orbit is obtained
 by an action that also changes $\psu_\mu$ in a non-trivial way. Also, the
 prefactor of $\tS$ gets changed analogously as in the formula in footnote
 \ref{futn}.}
\nxt
When $\calgcyd_a$ is non-trivial, then $\Abs a$ can no longer be written as
a linear combination of Ishibashi states of \CCY. This is simply due to
the fact that the preserved chiral symmetry is then not big enough
to guarantee that all \ACY-descendants are reflected at the boundary in
the same way, so that different descendants must be treated differently. The 
boundary state can, however, still be expressed in terms of suitable 
generalizations $\ishi{\cdots}$ of the Ishibashi states of the unextended 
theory \Cinner. These states are labeled by a pair $(\lambda,\psi_\lambda)$, 
where now again the restriction of zero monodromy charge is to be imposed on the 
primary field $\lambda$ (but no simple current orbit is taken any longer).\,%
 \futnote{Here $\psi_\lambda$ is always a character of the {\em full\/}
 stabilizer, even when the untwisted stabilizer is strictly smaller than
 the stabilizer \cite{fuSc11}.}

Heuristically the situation can be understood as follows. On the side of 
Ishibashi states, only states with vanishing monodromy charge are present; 
in the orbifold language of \cite{vafa4}, only states
in the untwisted sector appear. This means that the orbifold element in 
`space' direction on the torus is always trivial, whereas we still project,
i.e.\ we still deal with non-trivial group elements in `time' direction.
According to Cardy's ideas, the boundary states are obtained by a modular
S-transformation from the Ishibashi states. After that transformation we 
only have the trivial group element in `time' direction, which implies that 
there is no projection, so that orbits appear. However, in `space' direction 
we now have non-trivial elements, and therefore the twisted sectors show
up in the description of the boundary states.

To make these heuristic ideas quantitative, we introduce a 
matrix $\tS$ that takes over the role that the usual S-matrix plays in the 
Cardy case. As shown in \cite{fuSc11,fuSc12}, such a matrix indeed exists. The
following structures were uncovered.
\nxt
The expansion of the boundary state $\Abs a$ \wrt the generalized
Ishibashi states reads
  \be  \Abs{[\mu,\psi_\mu]} = \sum_{\lambda,\psi_\lambda}
  \frac{\tS_{(\lambda,\psi_\lambda),[\mu,\psi_\mu]}}
  {\sqrt{\tS_{(\vac),[\mu,\psi_\mu]}}} \, \ishi{(\lambda,\psi_\lambda)}
  \,.  \Labl77
The matrix $\tS$ appearing here can be expressed as
  \be  \tS_{(\lambda,\psi_\lambda),[\mu,\psi_\mu]}
  = \Frac{|\calg|}{|\cals_\lambda|\,|\cals_\mu|}
  \sum_{\J\in\cals_\lambda\cap\cals_\mu} \psi_\lambda(\J)\,
  \psi_\mu(\J)^*\, S^\J_{\lambda,\mu} \,,  \Labl tS
which is similar to \Erf xS (but remember that now also twisted sectors are 
allowed in the second label). Complete information on the boundary state, 
like brane tensions or RR charges, is encoded in this matrix $\tS$.
\nxt
Note the similarity between the result \Erf77 and Cardy's formula \cite{card9}
for symmetry preserving \bc s, in which the modular S-matrix appears
in place of $\tS$. It turns out that the matrix $\tS$ has still more in common 
with the modular S-matrix. Indeed, as first realized in \cite{prss2,prss3}, 
a subset of the sewing constraints \cite{lewe3} for \corfu s of a rational 
\cft\ can be isolated which leads to a simple non-linear equation for the 
bulk-boundary coefficients for excitation of the vacuum field on the boundary.
As pointed out in \cite{fuSc6}, this equation means that the
reflection coefficients constitute one-dimensional representations of a certain
\findim\ associative \alg, which generalizes
the fusion rule \alg\ and is called the {\em classifying \alg\/}.\,%
 \futnote{Extending
 Cardy's \cite{card9} work, it was shown in \cite{fuSc12} that the structure
 constants of this algebra are traces on suitable spaces of conformal blocks.}

These results allow us to introduce the notion of an {\em elementary
boundary condition\/}; this furnishes by definition an {\em irreducible\/} 
representation of the classifying algebra. Thus the elementary \bc s
are in one-to-one correspondence with the \onedim\ \irrep s of the classifying 
\alg. The matrix $\tS$ as given by \Erf tS diagonalizes the structure
constants of the classifying \alg, analogously as the modular S-matrix
diagonalizes the fusion rules. In string theory, on the other 
hand, one must in addition introduce (Chan\hy Paton) multiplicities 
for boundary conditions. Thus the space of all \bc s to be considered in
string theory forms a cone over the elementary \bc s, and
generically one is dealing with higher-dimensional, and hence necessarily 
reducible, representations of the classifying algebra.
It is also quite common that some of the solutions that are present as
elementary \bc s in the \cft\ possess a zero \CP multiplicity, i.e.\
do not appear at the string theory level at all.

Based on the properties of the matrix $\tS$, the space of boundary
conditions for \ACY\ that preserve \Ainner\ can be analyzed in detail
\cite{fuSc11,fuSc12}. One important result is that each of the 
\bs s studied here possesses a definite {\em \auto\ type\/}.
This means that it can be written as a linear combination of {\em twisted
Ishibashi states\/}, where the twist is a fixed automorphism $\omega$ of
\ACY. Such a twisted Ishibashi state $\ishio{[\lambda,\psi_\lambda]}$ obeys 
the twisted Ward identity
  \be
  \llb  \caly_n\oT\bfe + (-1)^{\Delta_\caly-1}\, \bfe\oT \omega(\caly_{-n})
  \lrb\, \ishio{[\lambda,\psi_\lambda]} = 0  \Labl9y
for every field $\caly$
of conformal weight $\Delta_\caly$ in \ACY. (This generalizes the usual 
definition of Ishibashi states of Dirichlet-type for the free boson.
In the terminology of \cite{kaok,reSC}, the relation \Erf9y says that 
$\omega$ provides the {\em glueing condition\/} of the \bc.)

The subset of \Ao-{\em type \bs s\/} -- that is, of all \bs s which 
preserve \Ainner\ and which have some prescribed \auto\ type $\omega$ --
corresponds to a sub\alg\ of the classifying \alg. The structure constants 
of this sub\alg\ can be expressed in terms of traces of the action of $\omega$
on the space of the relevant three-point conformal blocks. This again 
generalizes the Cardy case, in which the structure constants are just the 
fusion rules, which in turn are nothing else than traces of the identity 
automorphism on spaces of three-point blocks.

The fact that the boundary states that preserve \Ainner\ come in several 
different automorphism types can also be understood as follows. The identity 
map on \Ainner\ can be lifted to an automorphism of \ACY\ in several distinct
ways. The group of these lifts of automorphisms is just the quantum Galois 
group for the extension. Each element of this group gives us an automorphism 
type. Let us stress that the fact that all the boundary conditions that are
commonly referred to as A-type do possess an automorphism type is
non-trivial indeed. It reflects the equally non-trivial statement of quantum 
Galois theory that all intermediate algebras are obtained as fixed algebras.

Once the boundary states are given explicitly, i.e.\ once the matrix $\tS$
is known, all annulus amplitudes can be computed by sandwiching a string 
propagator $q^{L_0+\tilde L_0-c/12}$ between the two appropriate \bs s.
But already from the general expression \Erf tS above (and from
general properties of the matrices $S^\J$), it can be established via
representation theoretic arguments \cite{fuSc11}
that in full generality the annulus coefficients are non-negative integers,
as befits the coefficients of the open string partition function. 
The completeness and associativity properties \cite{prss3}
of the annulus coefficients can be shown to be satisfied as well.
Further, one can write the annulus amplitude for two \bc s of \auto\ types
$\omega_1$ and $\omega_2$ as a sum of characters $\chi'_{[\nu,\psi_\nu]'_{}}$
of the extension of \Cinner\ by the subgroup
  \be  \calgp := \{ \J\iN\calgcy \,|\, Q_\J(\omega_1)=0=Q_\J(\omega_2)\}  \ee
of $\calgcy$.
(Here the isomorphism between $\calgcyd$ and the dual group $(\calgcy)^*_{}$
is used, i.e.\ the automorphisms are regarded as characters of $\calgcy$.)
The coefficients then read \cite{fuSc11}\,%
 \futnote{Here again we suppress the changes that arise when genuine
 untwisted stabilizer groups are present; see formula (6.23) of \cite{fuSc11}.}
  \be  {\rm A}_{[\mu_1,\psi_1],[\mu_2,\psi_2]}^{[\nu,\psi_\nu]'_{}}
  = \Frac{|\calgp|}{|\calgcy|} \sum_{[\lambda,\psi_\lambda']'_{}}
  \Frac{|\cals_\lambda|}{|\cals_\lambda'|} \sum_{\psi_\lambda\succ\psi_\lambda'}
  \tS_{(\lambda,\psi_\lambda),[\mu_1,\psi_1]}^*
  \tS_{(\lambda,\psi_\lambda),[\mu_2,\psi_2]}^{}
  S'_{[\lambda,\psi_\lambda']_{}',[\nu,\psi_\nu]'_{}}
  / S'_{[\lambda,\psi_\lambda']'_{},[\vac]'_{}} \,,  \Labl'a
where $S'$ is the modular S-matrix of the $\calgp$-extension of \Cinner,
and where the second summation is over all $\cals_\lambda$-characters
$\psi_\lambda$ that restrict to the given character $\psi_\lambda'$ of the
subgroup $\cals'_\lambda\eq\cals_\lambda\,{\cap}\,\calgp$.

Another important conclusion to be drawn is that the monodromy charge
constitutes a grading of the annulus coefficients,
in the following sense. It follows from the result \Erf'a that 
  \be  {\rm A}_{[\mu_1,\psi_1],[\mu_2,\psi_2]}^{[\nu,\psi_\nu]'_{}}
  = \eE^{2\pi\ii[Q_\J(\mu_2)-Q_\J(\mu_1)+Q_\J(\nu)]}_{} \cdot
  {\rm A}_{[\mu_1,\psi_1],[\mu_2,\psi_2]}^{[\nu,\psi_\nu]'_{}} \,,  \ee
so that the annulus coefficient ${\rm A}_{[\mu_1,\psi_1],[\mu_2,\psi_2]}
^{[\nu,\psi_\nu]'_{}}$ vanishes unless
$Q_\J(\nu)\eq Q_\J(\mu_1)\,{-}\,Q_\J(\mu_2)$ for all $\J\iN\calgcy$. Thus 
all open string states that appear in the annulus amplitude for two boundary 
conditions of automorphism types $\omega_1$ and $\omega_2$
have a common monodromy charge $Q_\J$ \wrt any current $\J$ in $\calgcy$, and 
this common value is given by
  \be  \exp(2\pi\ii Q_\J) = \omega_1^{-1}\omega_2^{}(\J) \,.  \Labl12

\subsection{World sheet \susy\ and space-time \susy}\label{sss}

We now analyze what boundary conditions in string compactifications
of Gepner type preserve the (super)symmetries that have to be imposed 
to obtain a consistent bulk theory.
To this end we study in more detail the twisted Ward identities \Erf9y.
Again we start our discussion with boundary states that preserve the chiral 
symmetry \alg\ \Ainner\ of the inner sector \Cinner. As already mentioned, such 
boundary states are often collectively said to be of A-type, see for instance
\cite{reSC,oooy,staN3,gusa}. However, as we have detailed above, the inner
sector \Cinner\ of the string compactification needs to be extended to
\CCY, which does not have \Ainner\ as
chiral symmetry, but rather its simple current extension \ACY.
To identify boundary states that are relevant for string theory,
in particular those that can be given a geometric interpretation as
D-branes, it is necessary to understand what part of the extended 
algebra \ACY\ is preserved or broken by a given boundary state.

Recall that the simple current extension from \Cinner\ to \CCY\ consists
of a part that ensures world sheet supersymmetry and another part necessary
for space-time supersymmetry. 
Accordingly, the automorphism type $\omega_a$ of a boundary condition 
that preserves \Ainner\ carries information concerning both
the extension by the alignment currents $\ww_i\eq v_{1,\inner}v_{i,\inner}$ 
$(i\eq2,3,...\,,r$) and the extension by $\sv\eq s_{\inner}^2 v_{1,\inner}^{d/2}$. 
These individual pieces of information can be thought of as measuring which 
supersymmetries are broken or conserved by the boundary state. More concretely,
we can attribute to every boundary state $\Abs a$ an element $\omega_a$
of the orbifold group $\calgcyd$ that is dual to $\calgcy$. As an abstract abelian 
group, this is again $\zet_{N_s/2}{\times}\,\zet_2^{r-1-\eta}$.
The \auto\ $\omega_a$ occurs in the twisted Ward identities \Erf 9y 
satisfied by $\Abs a$; explicitly, we have
  \be  \bearll  \omega_a(\ww_i) = \zeta_{a,i}\, \ww_i & \ {\rm with}
  \quad \zeta_{a,i}\iN\{\pm1\} \,,  \\{} \\[-.6em]
  \omega_a(\sv) = \eE^{2\ii\vartheta_a}_{}\, \sv & \ {\rm with}
  \quad 2\vartheta_a\iN 2\pi\zet/(N_s/2) \,,  \eear \Labl31
and hence
  \be \bearl
  \llb (\ww_i)_n\oT\bfe + \zeta_{a,i}\,(-1)^{\Delta_{\ww_i}-1}\, \bfe\oT
  (\ww_i)_{-n} \lrb\, \Abs a = 0\,,  \\{} \\[-.5em]
  \llb \sv_n\oT\bfe + \eE^{2\ii\vartheta_a}_{}\, (-1)^{\Delta_\sv-1}\, \bfe\oT
  \sv_{-n} \lrb\, \Abs a = 0 \,.  \eear  \labl{gluco}

Also recall that for each tensor factor of the internal \nn theory
the field $v_{i,\inner}$ contains the \susy\ charges $G_i^\pm$.
Now $N\eq1$ {\em world sheet \susy\/} plays the role of a gauge symmetry of
perturbative superstring theory. In order not to destroy this constitutive
feature of superstrings, $N\eq1$ world sheet \susy\ must not be broken by any 
boundary state that is present (i.e., has a non-zero \CP multiplicity) at the 
string theory level. Concretely, we have the relation $\{\QB,\beta\}\eq G$ 
between the $N\eq1$ supercurrent $G$, the BRST charge $\QB$ and the 
superghost $\beta$. When combined with the identities
  \be  (\QB\oT\bfe - \bfe\oT \QB)\,\Abs a = 0  \qquad{\rm and}\qquad
  (\beta_r\oT\bfe + \ii\epsilon\, \bfe\oT\beta_{-r})\,\Abs a = 0  \ee
($\epsilon\iN\{\pm1\}$) which encode BRST invariance of the \bs\ and the \bc\
for the superghost (which is model independent and independent of the chosen 
\bc\ for the theory \CCY), this relation implies that we must also have
  \be  (G_r\oT\bfe + \ii\epsilon\, \bfe\oT G_{-r})\,\Abs a = 0 \,.  \ee
Comparison of this Ward identity with the result \erf{gluco} then tells us that 
we must require the invariance property $\omega_a(G)\eq G$
of the $N\eq1$ supercurrent.  

The extended theory \CCY\ actually possesses \nn world sheet \susy, with
supercurrents $G^\pm\eq\sum_iG_i^\pm$,
into which the $N\eq1$ \susy\ can be embedded in several different ways,
namely as $G\eq(\eE^{\ii\gamma}G^+{+}\,\eE^{-\ii\gamma}G^-)/\sqrt2$ for
any $\gamma\iN{\dl R}$. By inspecting the operator products of the 
supercurrents, it follows that there exists {\em some\/} $N\eq1$ sub\alg\ 
that is preserved if and only if in the relations \Erf31 we have $\zeta_{a,i}
\eq1$ for all $i$, i.e.\ if and only if the part of the automorphism type that 
concerns the alignment currents is trivial. (This appears to have been ignored
in part of the literature.) As a matter of fact, in that case the \bc\ 
preserves {\em all\/} $N\eq1$ sub\alg s, and hence even the whole \nn \alg.

Thus from now on we only admit those \auto\ types which obey
  \be  \omega_a(\ww_i) = \ww_i \,. \ee
(Put differently, in the string theory any \cft\ \bc\ which violates
this relation is assigned \CP multiplicity zero.) 
According to the Ward identities \erf{gluco},
the \auto\ type of any of the remaining \bc s is then completely 
characterized by a single number. This number is essentially given by
$\vartheta\iN 2\pi\zet/N_s$; but we also have to take into account
that the order $N_s'/2$ of the image $\sw$ of $\sv$ in the theory \Cwsusy\
can be different from the order $N_s/2$ of $\sv$ itself. Thus the \bc s
must rather be labeled by
  \be  \theta = 2\pi n/N_s' \quad{\rm with}\quad n\iN\{0,1,...\,,N_s'{-}1\}
  \,.  \ee
We shall denote the corresponding automorphism type of branes by A$_\theta$.
Note that here we use $\theta$ instead of $2\theta$ as label, even though 
from the point of view of the \CY extension, the automorphism types $\theta$ 
and $\theta{+}\pi$ cannot be distinguished. We do so because in full string 
theory (i.e., for \Cgepner\ rather than \CCY),
the two automorphism types differ on the final GSO projection.

This restriction to boundary conditions that preserve not only \Ainner,
but even its extension by the alignment currents i.e.\ the algebra \Awsusy, 
also ensures that the annulus partition function between any two boundary 
conditions has an open string spectrum with world sheet \susy.

In contrast to world sheet \susy, {\em space-time \susy\/} is 
not an indispensable ingredient of a superstring theory. Also, it is 
not determined by chiral considerations alone, in the sense that 
the space-time \susy\ generators $Q$ are given by closed string operators in
which left- and right-movers are combined.  Indeed, they arise as zero modes
  \be  Q = \int\! {\rm d}^2z\; s(z,\bar z)  \ee
   of fields in the full \cft\ obtained by putting together both chiral halves.
In an \nn string
compactification, the field $s(z,\bar z)$ consists of a spectral flow operator
on one chiral half combined with the vacuum field of the other chiral half
(for more explicit expressions, see e.g.\ \cite{Gepn,gusa}). 
Thus there are both left- and right-moving \susy\ charges, $Q_L$ and $Q_R$.
In the context of open strings, a preserved space-time 
\susy\ is a certain linear combination of a left- and right-moving 
charge that annihilates the boundary state \cite{polc7,Polc2}: 
  \be  (Q_L^{} + Q'_R)\, \Abs a = 0 \,.  \ee
If $s$ is any left-moving spectral flow operator, $\omega_a(s)$ will again 
have the properties of a spectral flow, and therefore the sum of the 
corresponding left and right-moving zero modes will still constitute a 
conserved \susy. That different boundary states in Gepner models conserve 
different space-time \susy\ charges has been observed in \cite{gusa}. 
What is new about the analysis above is the observation
that this corresponds to dealing with \bc s of distinct automorphism types.
Indeed, the fact that within the classification into A- and B-type there
exist subclasses of boundary states with different
automorphism type has so far not been appreciated in the literature.  

Thus all boundary conditions in Gepner models that 
have been considered so far in the literature, and for which we have 
identified an automorphism type labeled by an element $\theta\iN
2\pi\zet/N_s'$, in fact do preserve half of space-time \susy.
In short, in all situations studied in this paper,
the presence of {\em space-time\/} \susy\ in the closed string sector
(ensured by the GSO projection) together with the preservation of
$N\eq1$ {\em world sheet\/} \susy\ by a \bs\ already guarantees that 
the \bs\ is BPS. On the other hand, the open string spectrum, encoded 
in the annulus partition function, will be space-time supersymmetric 
only if the automorphism types of the boundary conditions on the two 
boundary components of the annulus are equal. (Without reference 
to automorphism types, a condition of this type for space-time 
\susy\ of the partition function was also derived in \cite{reSC}.)
In that case all annuli can be expressed in terms of characters of the
\CY extension \CCY, and as a consequence the GSO projection -- which is
a chiral issue -- guarantees in particular the absence of tachyons
in the spectrum of open string states. Geometrically, the difference 
$\theta_1\,{-}\,\theta_2$ between the labels of the two automorphism 
types can be given the intuitive interpretation of an angle between two 
branes. The open string spectrum is space-time \susic if the angle
between the two branes is zero. In contrast, when the angle is non-zero, then
absence of tachyons in the open string spectrum is not guaranteed any longer.

Thus in order to guarantee the absence of open string tachyons, among the 
A$_\theta$-type \bc s it is generically necessary to restrict to those with a 
single fixed value of $\theta$. For instance, when we decide to keep a \bs\ with
$\theta\eq0$, then we typically have to dismiss all \bs s with $\theta\,{\ne}\,0$. 
However, in the case of type I theories, an orientifold projection may stabilize 
the brane (for reviews see \cite{sen18,Leru}); in such circumstances, 
A$_\theta$ \bc s with several distinct values of $\theta$ could coexist.
Whether this happens or not, and for what choices of $\theta$, are model 
dependent questions. (The answer can in particular depend on bulk moduli.)
Furthermore, when we also include B-type conditions, it can happen that
requiring absence of open string tachyons restricts the allowed $\theta$ value
of A$_\theta$ conditions even when only a single $\theta$ is kept.

\subsection{\Ae-type \bc s}

The special class of \bc s of type $\Ae\,{\equiv}\,{\rm A}_{\theta=0}$
are just those of `trivial automorphism type' $\omega\eq\id$, i.e.\ those for 
which the identity map of \ACY\ is used as the extension of the identity of 
\Ainner. These \bc s preserve the whole 
algebra \ACY. Put differently, they are precisely the \bc s that were studied
long ago by Cardy \cite{card9} for an arbitrary rational \cft. 
In the case of tensor products of \mimo s, various aspects of 
\Ae-type conditions have already been studied in \cite{reSC,gusa,gojs,brSc}.
As far as the space-time aspects of string theory are concerned, the \Ae-type
\bc s are not distinguished in any specific manner among the larger set of
A-type \bc s, which as explained above preserve both world sheet and space-time
\susy, too. However, from a pure world sheet point of view, \Ae-type \bc s
are special in that they preserve the full chiral algebra \ACY\ of the
\CY extension. This means that they are directly accessibly by Cardy's method, 
and we therefore still treat them separately here.

The only data that enter Cardy's construction of \bs s are the entries of the 
modular S-matrix of the \cft, i.e.\ in our case the matrix \Erf xS for the \CY 
extension \CCY. (In terms of the classifying algebra mentioned above, this result 
is a manifestation of the fact that the invariant subalgebra that corresponds to 
the \Ae-type \bc s is nothing but the fusion rule algebra of \CCY.) These data are 
well under control (some explicit formulas will be presented below). Besides 
the \bs s, also other aspects of these special \bc s are well understood 
(compare e.g.\ \cite{fffs2,fffs3} for a general discussion of correlation
functions). Therefore the case of trivial automorphism type -- which was also 
used as a starting point in the constructions in \cite{fuSc11,fuSc12} -- is
absolutely under control. 

In particular, the \Ae-type \bs s are in natural
one-to-one correspondence with the primary fields of the \CY extension
\CCY, and they can be expanded in the Ishibashi states of \CCY:
  \be  \Aebs{[\mu,\psi_\mu]} = \sum_{[\lambda,\psi_\lambda]}
  \frac{\Scy_{[\lambda,\psi_\lambda],[\mu,\psi_\mu]}}
  {\sqrt{\Scy_{[\vac],[\mu,\psi_\mu]}}} \, \ishi{[\lambda,\psi_\lambda]} \,.  
  \labl{Aebs}
Similarly, in the formula \Erf'a for the annulus coefficients we now have 
$\calgp\eq\calgcy$ as well as $\tS\eq S'\eq S\cy$, so that it reduces to
  \be  {\rm A}_{[\mu_1,\psi_1],[\mu_2,\psi_2]}^{[\nu,\psi_\nu]}
  = \sum_{[\lambda,\psi_\lambda]}
  \Scy_{[\lambda,\psi_\lambda],[\mu_1,\psi_1]}^*
  \Scy_{[\lambda,\psi_\lambda],[\mu_2,\psi_2]}^{}
  \Scy_{[\lambda,\psi_\lambda],[\nu,\psi_\nu]}
  / \Scy_{[\lambda,\psi_\lambda],[\vac]} \,.  \ee
By comparison with the Verlinde formula for \CCY\ we then learn that the annulus 
coefficients indeed coincide with the structure constants of the fusion \alg\ of
\CCY.

\subsection{Tensor products of \mimo s} \label{ssmm}

Let us now specialize to the \gemo s proper, where the inner sector \Cinner\
is a tensor product $\Ctensor\eq{\cal C}_{k_1}{\otimes}\cdots{\otimes}\,
{\cal C}_{k_r}$ of \nn minimal models ${\cal C}_{k_i}$ at levels 
$k_i\iN\zetplus$. We also restrict our attention to the \Ae-type \bc s. 
In this special case it is particularly easy to make the formula 
for $\tS$, and hence the description of \bs s, fully explicit. We first recall
the following information about such \gemo s that will be needed in the sequel.
\nxt
It ist convenient to think of an \nnmimo\ at level $k$ as a coset construction
$\sutwo_k{\times}\,\uone_4/\uone_{2h}$, where $h\,{:=}\,k{+}2$.  Accordingly, 
we denote its primary fields by $\Phi^{l,s}_q$. Then the labels $l,s,q$
are in the ranges $l\iN\{0,1,...\,,k\}$, $s\iN\{0,1,2,3\}$, and
$q\iN\{0,1,...\,,2h{-}1\}$, subject to the parity selection rule 
$l+s-q\iN2\zet$, and to the `field identification' \cite{gepn8,scya4}
$\Phi^{l,s}_q\,{\equiv}\,\Phi^{k-l,s+2}_{q+h}$. This labeling of primary
fields refers to the bosonic subalgebra of the \nn superconformal
algebra. For example, the two world sheet supercurrents are not
regarded as descendants of the vacuum but, rather, both correspond to the 
primary field $\Phi^{0,2}_0\,{\equiv}\,\Phi^{k,0}_{\pm h}$.
The identity primary field is $\vac\eq\Phi^{0,0}_0$.
(For more details about \mimo s see, for example, \cite{scya4}.)
\nxt
The primary fields of \Ctensor\ are then labeled as
   \be  (\Phi^{l_1,s_1}_{q_1}\,, \Phi^{l_2,s_2}_{q_2}\,,...
   \,, \Phi^{l_r,s_r}_{q_r} )  \Labl2r
where $\Phi^{l_i,s_i}_{q_i}$ is a primary field of the $i$th \mimo.
For brevity, below we will also use the notation
  \be  (\lambda,\sigma,\xi)\equiv(l_1,s_1,q_1,...\,,l_r,s_r,q_r)  \ee
for these collections of labels.
\nxt
The simple current group $\calgcy$ of the \CY extension is of the form \erf{gcy},
with $r$ the number of \mimo\ factors. It is generated 
by the $r{-}1$ order-two currents
  \be  \ww_i := (v,\vac,...\,,\vac,v,\vac,...\,,\vac)\,,
  \qquad i\iN\{2,3,...\,,r\} \,,  \Labl2v
where the $v$-entries are in the first (say) and $i$th \mimo,
together with the combination
  \be  \sv := (s,s,...\,,s)^2_{}\,(v,\vac,...\,,\vac)^{d/2}_{}
  = (s^2v^{d/2},s^2,...\,,s^2) \,. \Labl sv
Here $v$ stands for the \mimo\ primary field $\Phi^{0,2}_0$ that contains the
two world sheet supercurrents, and $s\eq\Phi^{0,1}_1$ is the simple current
in the Ramond sector whose action provides the spectral flow. These fields
  \be  s=\Phi^{0,1}_1  \ \quad{\rm and}\quad\ v=\Phi^{0,2}_0  \Labl3s
have conformal weight $c/24\eq k/8(k{+}2)$ and $3/2$, respectively.
\nxt
The order of $\sv$ is
  \be  {\rm ord}(\sv) = N_s/2 = {\rm scm}_{i=1,2...r} \{ \eta_i\,h_i \}
  \Labl2n
with $h_i\,{\equiv}\,k_i{+}2$ and $\eta_i\eq1$ for $k_i$ even,
$\eta_i\eq2$ for $k_i$ odd.

Exploiting our knowledge about the \mimo\ fusion rules, 
it is straightforward combinatorics to establish the following
group and fixed point structure of $\calgcy$.
\nxt
When all levels $k_i$ are odd, then we have
  \be  \sv^{N_s/4} = \prod_{i=1}^r \ww_i \,,  \Labl2o
and hence $\calgcy\eq\zet_{N_s/2}\Times \zet_2^{r-2}$.
In this case there are no fixed points.
\nxt
In contrast, when at least one level is even, then we have 
$\calgcy\eq\zet_{N_s/2}\Times\zet_2^{r-1}$, and fixed points do occur. 
In fact there is then a unique simple current
$\L\iN\calgcy$ having fixed points. $\L$ is given by
  \be  \L = \sv^{N_s/4}_{} \prod_{i=1}^r (\ww_i)^{\eps_i} \,, \labl L
where the value of $\epsilon_i\iN\{0,1\}$ depends on the power of 2 contained
in $N_s/2$. Namely, according to \Erf2n, $N_s/2$ is always even; when it is also
divisible by 4, then
  \be  \eps_i = \left\{ \bearll
  1 & \mbox{when the power of 2 in $h_i$ is maximal}\,, \\[.1em]
  0 & {\rm else}\,.  \eear\right. \ee
When $N_s/2$ is not divisible by 4, then we have instead
  \be  \eps_i = \left\{ \bearll
  1 & \mbox{when $h_i$ is odd}\,, \\[.1em]
  0 & {\rm else}\,.  \eear\right. \ee

It also follows that $\L$ has order 2, and that with the help of field
identification it can be rewritten as
  \be  \L = (\Vac,...\,,\Vac,\Phi^{k_{r'+1},0}_0,...\,,\Phi^{k_r,0}_0) \,.
  \Labl LL
Here without loss of generality we assume that the $h_i$ have been ordered
in such a way that those containing the maximal power of 2 are the last
$r{-}r'$ ones, i.e.\ are labeled by $i\iN\{r'{+}1,...\,,r\}$. Fixed points
of $\L$ are all fields \Erf2r which obey 
$l_i\eq k_i/2$ for every $i\eq r'{+}1,...\,,r$, while all other labels are 
arbitrary (except, of course, for the parity selection rule $l_i{+}s_i{-}
q_i\iN2\zet$ and the restriction to zero monodromy charge \wrt $\calgcy$).

Employing the general formula \Erf xS, we are thus in a position to display
the modular S-matrix of the \CY extension. For the tensor product theory 
\Ctensor\ we have the tensor product of the S-matrices of the individual 
factors:
  \be  S_{(\lambda,\sigma,\xi),(\lambda',\sigma',\xi')} = 2^r \, 
  \prod_{i=1}^r S^{\sutwo_{k_i}}_{l_i^{},l_i'}\, S^{\uone_4}_{s_i^{},s_i'}\,
  \llb S^{\uone_{2h_i}}_{q_i^{},q_i'} \lrb^*  \ee
(the $r$ factors of $2$ stem from field identification in each of the $r$
minimal models). The fixed point matrix $S^{\L}$ reads
  \be  S^{\L}_{(\lambda,\sigma,\xi)^\L,(\lambda',\sigma',\xi')^\L}
  = 2^r\, \prod_{i=1}^r S^{\uone_4}_{s_i^{},s_i'}\,
  \bigl(S^{\uone_{2h_i}}_{q_i^{},q_i'}\bigr)^*
  \prod_{i=1}^{r'} S^{\sutwo_{k_i}}_{l_i^{},l_i'}
  \prod_{i=r'+1}^r S^{\J\,\sutwo_{k_i}}_{k_i/2,k_i/2} \,. \ee
(This is only defined when both fields involved are fixed points,
which is indicated by attaching the superscript of the labels $(\cdots)^\L$.)
Here $S^{\J\,\sutwo_k}$ is the fixed point `matrix' for the simple
current $l\eq k$ of the $\sutwo$ \wzwm\ at level $k$. As $k$ acts by fusion
as $k\,{\star}\,l\eq k{-}l$ and hence has just a single fixed point $k/2$, 
the quantity $S^{\J\,\sutwo_k}$ is in fact a single number, and in this
simple situation the results of \cite{scya6} tell us immediately that
$S^{\J\,\sutwo_k}_{k/2,k/2}\eq\eE^{-2\pi\ii\cdot 3k/16}$. Moreover, 
$\sum_{i=r'+1}^rk_i/16$ must be a multiple of $1/12$ \cite{scya6}, so that we 
have $\prod_{i=r'+1}^r S^{\J\,\sutwo_{k_i}}_{k_i/2,k_i/2}\eq\eE^{\pi\ii\,p/2}$
with $p\iN\{0,1,2,3\}$ (inspecting the list of \gemo s, one sees that
all values of $p$ occur). For further simplification, notice that
  \be  \prod_{i=r'+1}^r (S^{\J\,\sutwo_{k_i}})^2_{}
  = (-1)^{\sum_{i=r'+1}^r 3k_i/4} = (-1)^{\sum_{i=r'+1}^r k_i/4}  \Labl26
and that $\sum_{i=r'+1}^r k_i/4$ is always an integer, 
as follows from the condition on the central charge.

Putting this information together, we see that the modular S-matrix of \CCY\ reads
  \be \bearl
  S\cy_{[(\lambda,\sigma,\xi),\psi],[(\lambda',\sigma',\xi'),\psi']}
  = 2^{2r-2-\eta}N_s\, \dsty\prod_{i=1}^r S^{\uone_4}_{s_i^{},s_i'}\,
  \bigl(S^{\uone_{2h_i}}_{q_i^{},q_i'}\bigr)^*
  \prod_{i=1}^{r'} S^{\sutwo_{k_i}}_{l_i^{},l_i'}
      \\{}\\[-1.05em] \hsp{3.5}
  \Llb \dsty
  \llb 1-\Frac34\prod_{i=r'+1}^r\!\!\delta_{l_i^{},k_i^{}/2}\delta_{l_i',k_i^{}/2} \lrb
  \prod_{i=r'+1}^r S^{\sutwo_{k_i}}_{l_i^{},l_i'}
  + \Frac14\, \psi\psi'\!\!\prod_{i=r'+1}^r
  \delta_{l_i^{},k_i^{}/2}\delta_{l_i',k_i^{}/2}\, S^{\J\,\sutwo_{k_i}}_{k_i/2,k_i/2}
  \Lrb
      \\{}\\[-.7em] \hsp{2.2}
  = \Frac{2^{r-2-\eta}\,N_s}{\prod_{i=1}^r h_i}\,
  \dsty\prod_{i=1}^r \eE^{\pi\ii (q_i^{}q_i'/h_i^{} - s_i^{}s_i'/2)}
  \prod_{i=1}^{r'} \sin\llb \Frac{(l_i^{}+1)(l_i'+1)\pi}{h_i} \lrb
      \\{}\\[-1.15em] \hsp{3.5}
  \Llb \dsty
  \llb 1-\Frac34\prod_{i=r'+1}^r\!\!\delta_{l_i^{},k_i^{}/2}\delta_{l_i',k_i^{}/2} \lrb
  \prod_{i=r'+1}^r\! \sin\llb \Frac{(l_i^{}+1)(l_i'+1)\pi}{h_i} \lrb
      \\{}\\[-1.3em] \hsp{11.7}
  + 2^{-2+(r-r')/2}\, (-1)^{\sum_{i=r'+1}^r k_i/4}\, \psi\psi'\!\!  \dsty
  \prod_{i=r'+1}^r\! \delta_{l_i^{},k_i^{}/2}\delta_{l_i',k_i^{}/2}\, h_i^{1/2} \Lrb
  \,. \eear  \labl{Scy}
Recall that $\eta\eq1$ when all levels are odd, in which case there are no fixed
points, and $\eta\eq0$ else. Upon insertion of \erf{Scy} into the Verlinde 
formula, one obtains the fusion rule coefficients of \CCY; for $\eta\eq0$ we 
arrive at the following expression (for $\eta\eq1$ the formula is similar, but 
without the complications involving fixed points):
  \be \bearl
  \Ncy{[(\lambda,\sigma,\xi),\psi]}{[(\lambda',\sigma',\xi'),\psi']}
  {[(\lambda'',\sigma'',\xi''),\psi'']}
  = \Frac1{|\cals_\lambda^{}|\,|\cals_{\lambda'}|\,|\cals_{\lambda''}|}\,
  \dsty\sum_{n=0,...,N_s\!/4-1}
  \sum_{\scs\eps_j=0,1\;\forall\;j=1,2...,r \atop \sum_j\eps_j=0 \bmod 2}\!\!
      \\{}\\[-1.5em] \hsp1
  \dsty\prod_{i=1}^{r'}
  \llb \delta_{\ds_i+2n+2\eps_i}\delta_{\dq_i+2n}\, \N {l_i^{}}{l_i'}{l_i''}
     + \delta_{\ds_i+2n+2\eps_i+2}\delta_{\dq_i+2n+h_i}\,
       \N {l_i^{}}{k_i^{}-l_i'}{\ \ \ l_i''}
  \lrb
      \\{}\\[-.8em] \hsp{1.5}
  \LLb \prod_{i=r'+1}^r
  \llb \delta_{\ds_i+2n+2\eps_i}\delta_{\dq_i+2n}\, \N {l_i^{}}{l_i'}{l_i''}
     + \delta_{\ds_i+2n+2\eps_i+2}\delta_{\dq_i+2n+h_i}\,
       \N {l_i^{}}{k_i^{}-l_i'}{\ \ \ l_i''}
  \lrb
      \\{}\\[-.6em] \hsp3
  +  \prod_{i=r'+1}^r
  \llb \delta_{\ds_i+2n+2\eps_i}\delta_{\dq_i+2n}\,
       \N {l_i^{}}{k_i^{}-l_i'}{\ \ \ l_i''}
     + \delta_{\ds_i+2n+2\eps_i+2}\delta_{\dq_i+2n+h_i}\,\N{l_i^{}}{l_i'}{l_i''}
  \lrb
      \\{}\\[-.6em] \hsp3
  + 2\,\prod_{i=r'+1}^r \delta_{\ds_i+2n+2\eps_i}\delta_{\dq_i+2n}\,
  \Llb \psi \psi'  \prod_{i=r'+1}^r (S^{\J\,\sutwo_{k_i}})^2_{}\,
       \delta_{l_i^{},k_i^{}/2}\delta_{l_i',k_i^{}/2}
       \Frac{S^*_{l_i'',k_i^{}/2}}{S_{0,k_i/2}}
      \\{}\\[-.6em] \hsp7
     + \psi'\psi'' \prod_{i=r'+1}^r
       \delta_{l_i',k_i^{}/2}\delta_{l_i'',k_i^{}/2}
       \Frac{S_{l_i^{},k_i^{}/2}}{S_{0,k_i/2}}
     + \psi \psi'' \prod_{i=r'+1}^r
       \delta_{l_i^{},k_i^{}/2}\delta_{l_i'',k_i^{}/2}
       \Frac{S_{l_i',k_i^{}/2}}{S_{0,k_i/2}}
  \Lrb
      \\{}\\[-.6em] \hsp3
  + 2\,\prod_{i=r'+1}^r \delta_{\ds_i+2n+2\eps_i+2}\delta_{\dq_i+2n+h_i}\,
  \Llb \psi \psi'  \prod_{i=r'+1}^r (S^{\J\,\sutwo_{k_i}})^2_{}\,
       \delta_{l_i^{},k_i^{}/2}\delta_{l_i',k_i^{}/2}
       \Frac{S^*_{k_i^{}-l_i'',k_i^{}/2}}{S_{0,k_i/2}}
      \\{}\\[-.6em] \hsp7
     + \psi'\psi'' \prod_{i=r'+1}^r
       \delta_{l_i',k_i^{}/2}\delta_{l_i'',k_i^{}/2}
       \Frac{S_{k_i^{}-l_i^{},k_i/2}}{S_{0,k_i/2}}
     + \psi \psi'' \prod_{i=r'+1}^r
       \delta_{l_i^{},k_i^{}/2}\delta_{l_i'',k_i^{}/2}
       \Frac{S_{k_i^{}-l_i',k_i^{}/2}}{S_{0,k_i/2}}
  \Lrb \LRb \,.
  \eear \labl{Ncy}
Concerning the notation, the following remarks are in order. First, we put
  \be  
  \ds_i^{}:=s_i^{}+s_i'-s_i'' \,,\qquad \dq_i^{}:=q_i^{}+q_i'-q_i'' \,.  \ee
Second, the factors $\delta_x\,{\equiv}\,\delta_{x,0}$ represent the fusion
coefficients of the various $\uone$ factors; in particular, an appropriate 
periodicity in their subscripts is understood. Third, we have separated the 
$\calgcy$-summation into a part involving the simple currents without fixed 
points ($\sum_{n,\eps_i}$) and one that implements the order-two simple current
$\L$ (the expression in curly brackets), compare formula \erf{z2fp}. Further, 
the innermost (pairwise) summation takes into account the field identification 
in the \mimo s.  Finally, $\psi\,{\equiv}\,\psi(\L)\iN\{\pm1\}$ corresponds
to the two irreducible characters of the group $\zet_2\eq\{\vac,\L\}$.

The terms in formula \erf{Ncy} that involve factors of $\psi$ can be simplified
by using the identity \Erf26 and noting that
  \be  \frac{S_{l,k/2}}{S_{0,k/2}} = \sin\llb\Frac{\pi}{2}\,(l\,{+}\,1)\lrb
  = \left\{\begin{array}{rl} (-1)^{l/2} & \mbox{if $l$ is even}\,, \\
  0 & \mbox{if $l$ is odd}\,.  \end{array}\right.  \ee
Both of the expressions in square brackets are thus non-zero only if the labels
$l_i^{},\,l_i',\,l_i''$ are even for all $i\iN\{r'{+}1,r'{+}2,...\,,r\}$, in 
which case they read
  \be  \psi \psi'\, (-1)^{\sum_{i=r'+1}^r k_i/4 + l_i''/2}\, \Pi\,\Pi'
  +\psi'\psi''\, (-1)^{\sum_{i=r'+1}^r l_i/2}\, \Pi'\,\Pi''
  +\psi \psi''\, (-1)^{\sum_{i=r'+1}^r l_i'/2}\, \Pi\,\Pi'' \,,   \ee
where we introduced $\Pi\,{:=}\prod_{i=r'+1}^r\delta_{l_i,k_i/2}$ and 
analogously $\Pi'$ and $\Pi''$. We also mention that the annulus coefficients 
for annuli with two boundary conditions of \Ae-type are just given by the 
fusion rules, so that the expression \erf{Ncy} directly provides us with 
the multiplicities of the corresponding open string states.  

\subsection{\Ae-type \bc s for \mimo\ tensor products}

Having collected these results about \mimo\ tensor products, we are now in a
position to write down the \Ae-type \bs s for the corresponding
\CY extensions \CCY. According to Cardy's results, the labeling of the \Ae-type 
\bc s is {\em precisely the same\/} as for the primary fields of \CCY. Let us make 
this explicit. We start with the collection $(\lambda,\sigma,\xi)\,{\equiv}\,
(l_1,s_1,q_1,...\,,l_r,s_r,q_r)$ of labels ranging over $l_i\iN\{0,1,...\,,k_i\}$,
$s_i\iN\{0,1,2,3\}$, and $q_i\iN\{0,1,...\,,2h_i\}$. 
We then implement the various projections by 
imposing the following selections and identifications (both on
bulk fields and on boundary conditions).
\nxt
Selections: 
\\
We impose $l_i\,{+}\,s_i\,{+}\,q_i\iN2\zet$ (minimal model selection rule),
$s_1\,{+}\,s_i\iN2\zet$ for all $i\eq2,3,...\,,r$ (fermion alignment), and
$Q\,{\equiv}\,\sum_{i=1}^r (-q_i/h_i + s_i/2) - s_1 d/4
\iN\zet$ (charge projection).
\nxt
Identifications:
\\
We restrict the $l_i$ to the range $0\le l_i\le k_i/2$ (minimal model field
identification). Representatives for the orbits \wrt the alignment 
currents $\ww_i$ can be labeled
by a single $s\iN\{0,1\}$ when $d/2\,{+}\,r$ is odd, and $s\iN\{0,1,2,3\}$ when 
$d/2\,{+}\,r$ is even; all the $s_i$ are equal to $s$ $\bmod\, 2$.
\\
Implementation of the identification implied by the current $\sv$ is more 
difficult; it involves the divisibility properties of the heights $h_i$, 
which in general do not have a simple structure. In special cases, for 
instance when ${\rm lcm}_i\{h_i\}\eq h_j$ for some $j$, the corresponding
label $q_j$ can be set to zero using this identification.

Explicit formulas for \bs s of \gemo s were first presented in \cite{reSC},
where the \bs s were expressed in terms of the modular 
S-matrices of the $\sutwo$ and $\uone$ building blocks of the minimal models.   
However, as explained above, the chiral algebra \ACY\ of the Gepner model is
much larger than the chiral algebra \Atensor\ of the tensor product
\Ctensor\ of \mimo s. Accordingly, in the bulk the modular transformations 
are described by the extended S-matrix $S\cy$. The non-trivial
information contained in $S\cy$ that cannot be obtained from the
tensor product S-matrix alone stems from the presence of fixed points under 
the \CY extension. Once $S\cy$ has been determined,
the usual description of \bc s that preserve the full chiral algebra
\cite{card9,card10} can be applied, though it is now to be formulated 
with the help of the matrix $S\cy$ rather than $S$. It is therefore not 
guaranteed that the \bs s can be written in the form presented in \cite{reSC}.
That the modular S-matrix must be `properly resolved' for the construction 
of \bs s in \gemo s was emphasized in \cite{gojs}. It was also noticed 
in \cite{gojs} that some of the boundary states given in \cite{reSC} 
are not consistent with all projections in Gepner models, which explains 
certain discrepancies between the results of \cite{gusa} and \cite{gojs}.
More recently, it has been established in
\cite{brSc} that some of the \bs s of \cite{reSC} are not elementary.

In \cite{brSc} A-type \bs s for Gepner models with $D\eq4$ and $r\eq5$
were constructed by implementing the simple current extension (including in
particular fixed point resolution) directly on the \bs s of the underlying
\nn tensor product. On the other hand, the results reported above clearly 
also allow to obtain these A-type \bs s by performing the extension already 
in the {\em bulk\/}. 
When doing so, one obtains the \bs s by merely combining standard results
for simple current extensions in the bulk (which lead in particular to
the formula \erf{Scy} for the modular S-matrix $S\cy$ of the \CY extension)
with the general results of Cardy \cite{card9} for symmetry preserving
boundary conditions in arbitrary rational \cfts.
Let us point out that this way we get the A-type boundary states
for {\em every\/} \gemo, i.e.\ for $D\eq4,6,8$ and for any allowed $r$.\,%
 \futnote{To make contact to geometry, $D/2{+}r$ must be odd. This can,
 however, always be achieved by introducing a trivial $k\eq0$ \mimo\ as an
 additional factor of the tensor product.}
Moreover, when proceeding in this manner the \bs s are completely determined,
up to over-all normalization, by the algorithm. Thus there is e.g.\ no need
to invoke integrality of the annulus coefficients in order to fix the 
relative strength of the contributions from twisted and untwisted sectors. 
Indeed, integrality of the annulus coefficients involving only
\Ae-type \bc s coefficients is just the Verlinde formula
as applied to the extended theory, i.e.\ to the \CY extension.

More concretely, the \Ae-type boundary states are given by the expression
\erf{Aebs}, 
  \be
  \Aebs{[(\lambda,\sigma,\xi),\psi]} = \sum_{[(\lambda',\sigma',\xi'),\psi']}
  \frac{\Scy_{[(\lambda,\sigma,\xi),\psi],[(\lambda',\sigma',\xi'),\psi']}}
  {\sqrt{\Scy_{[0,0,0],[\lambda',\sigma',\xi'),\psi']}}} \,
  \ishi{[(\lambda',\sigma',\xi'),\psi']}  \labl{bsaid} 
with $S\cy$ as presented in formula \erf{Scy} and with the summation
ranging over all primaries of \CCY,\,%
 \futnote{We may wish to split the boundary state $\Abs a$ into its 
 contributions involving only the Ishibashi states for full $\calgcy$-orbits
 and only those for fixed points, respectively. Then the number
 $\prod_{i=r'+1}^r h_i^{1/2}$ that is built in $S\cy$ (see the last line of
 formula \erf{Scy}) appears as a factor between the two parts. In \cite{brSc}
 this factor was obtained     
 by imposing integrality of annulus coefficients.\label{$h_i^{-1/2}$}}
and the annulus coefficients 
for two \bc s of \Ae-type coincide with the fusion rules \erf{Ncy}.
In the special case that $(\lambda,\sigma,\xi)$ is {\em not\/} a fixed point
of the current $\L$ \erf L, the expression \erf{bsaid} for the \bs\ reduces to
  \be  \bearll  \Aebs{[(\lambda,\sigma,\xi)]} \!\!
  &= \Llb\Frac{2^{r-2-\eta}N_s}{\prod_{j=1}^r h_j}\Lrb^{1/2} 
  \!\dsty\sum_{[(\lambda',\sigma',\xi'),\psi']}
  \\{}\\[-.8em] & \hsp{2.4}
  \!\dsty\prod_{i=1}^r \eE^{\pi\ii (q_i^{}q_i'/h_i^{} - s_i^{}s_i'/2)}\,
  \frac {\sin\!\llb (l_i^{}{+}1)(l_i'{+}1)\pi/h_i^{} \lrb}
  {\llb \sin\!\llb (l_i'{+}1)\pi/h_i^{} \lrb\lrb_{}^{1/2}}\, 
  \ishi{[(\lambda',\sigma',\xi'),\psi']} \,.  \eear \labl{Aebs0} 
For the boundary state in the full Gepner model, one has to combine 
this expression with the space-time part of the boundary state,
include a phase factor $S^\exf_{\st,\st'}/\sqrt{S^\exf_{o,\st'}}$,
a sign from undoing the bosonic string map, and a factor of $2$
from the remaining projections. The resulting formula is then essentially
the one reported in \cite{reSC}; but even in this special case our result
differs in the power of 2 that appears in the prefactor.

In contrast, when $(\lambda,\sigma,\xi)$ {\em is\/} a fixed point of $\L$,
then the additional terms in the formula \erf{Scy} for $S\cy$ contribute,
and the expression for the \bs\ gets a bit lengthier. Boundary states 
$\Aebs{[(\lambda,\sigma,\xi),\psi]}$ corresponding to resolved fixed points 
have been studied in \cite{brSc}. Apart from the proper power of 2, our result 
differs from the one presented in \cite{brSc} also by the absence of a factor 
of $\sqrt{N_s}$ in the $\psi$-dependent terms.

\subsection{General A-type boundary conditions}

Boundary conditions of \auto\ type A$_\theta$ with $\theta\,{\ne}\,0$
cannot be obtained from Cardy's results alone. In this subsection, we explain
more explicitly how the generalizations developed in \cite{fuSc11,fuSc12} 
allow to structurize and construct all A-type boundary conditions.  
We are faced with the following situation. Given the theory \CCY, obtained
from \Cinner\ by a simple current extension with $\calgcy$, we
want to construct all boundary conditions that preserve at least 
the chiral symmetry \alg\ of \Cwsusy. Here \Cwsusy\ is the theory obtained 
from \Cinner\ by extension with the alignment currents $\ww_i$ alone; this 
means in particular that all \bc s preserve world sheet \susy.

The solution to this problem is as follows. The set of all such boundary 
conditions is the set of irreducible representations of the classifying 
algebra $\cla\equiv\cla(\CCY{:}\,\Cwsusy)$.  This set of \bc s can be 
divided into subsets of definite automorphism type, here labeled by the angle
$2\theta$ with $\theta/2\pi \iN\zet/N_s'$. (Recall from subsection \ref{sss}
that $\theta$ can be interpreted as specifying which space-time \susy\ 
is preserved by a \bc.) Furthermore, each of these subsets furnishes
the set of \irrep s of an {\em individual classifying algebra\/} 
$\cla_{2\theta}$ which, just like \cla, is a semisimple commutative 
associative \alg\ over $\dl C$\,, and one has
  \be \cla = \bigoplus_{\theta} \cla_{2\theta}  \ee
as a direct sum of \alg s over $\dl C$\,. 
We remark that the same situation arises in much simpler models in 
statistical mechanics, too. In the critical three-state Potts model, 
for instance, there are eight boundary conditions which come in two 
automorphism types that are distinguished 
by the reflection condition for the $W_3$-current \cite{fuSc9}.
Accordingly, in that case the classifying algebra is a direct sum
$\cla_+{\oplus}\,\cla_-$ of a six- and a two-dimensional algebra;  
the \irrep s of $\cla_+$ provide the three fixed and the three mixed
\bc s of the Potts model, while the ones of $\cla_-$ provide
the free and `new' \cite{afos} conditions.

In general there is no simple relationship between the various individual
classifying algebras $\cla_{2\theta}$. In the case of 
Gepner models, however, it turns out that there exist symmetries
generated by simple currents of the theory \Cwsusy\ (the so-called `phase
symmetries'), which act on the set of boundary conditions and thereby relate
individual classifying algebras for different $\theta$ with each other. 
Because of those symmetries, the \bc s for $\theta\,{\ne}\,0$ look
still very similar to the Cardy case. It follows in 
particular that the combinatorics of fixed points and their resolution 
do not depend on the value of $\theta$. (This can already be deduced from 
formula \Erf LL for the fixed point current $\L$.) Those symmetries between 
boundary conditions with different $\theta$ have been implicitly used in 
\cite{bDlr} for a nice organization of A-type \bc s. Note, however, that
this simplification is intrinsically linked to the special 
symmetries of \nnmimo s and cannot be expected to be present in general.

For a more detailed description, we recall that the group that furnishes 
the extension from \Cwsusy\ to \CCY\ is generated by the image $\sw$ 
of the simple current $\sv$ \Erf sv in the \Cwsusy-theory, and that this
current has order $N_s'/2$ with $N_s'$ as defined in formula \Erf uu,
i.e.\  $N_s'\eq N_s/2^\eta$ with $\eta\iN\{0,1\}$. The primary 
fields of \Cwsusy\ are labeled by $\Lambda\,{:=}\,[(\lambda,\sigma,\xi)]'$
where the prime indicates that orbits are taken \wrtt group generated by the
alignment currents only (which do not have fixed points); thus in particular
$\sw\eq[\sv]'$. A distinguished basis $\{\tilde\Phi_\Lambda\}$ of the 
classifying algebra $\cla$ is then labeled by the set of all \Cwsusy-fields 
$\Lambda$ that have vanishing monodromy charge \wrt $\sw$, together with a 
$\zet_2$ character accounting for a possibly non-trivial stabilizer. Further,
natural bases of the {\em individual\/} classifying algebras $\cla_{2\theta}$ 
are provided by twisted sums $\tilde\Phi^\theta_\Lambda\,{:=} \sum_{j=0}
^{(N_s'/2)-1}\eE^{2\ii j\,\theta}\!\tilde\Phi_{\sw^j_{\phantom|}\Lambda}$ 
of the basis elements $\tilde\Phi_\Lambda$ of $\cla$.  
The set of all \bc s, on the other hand, is labeled by the set of {\em orbits\/}
$[\Lambda,\psi]''$ of fields of \Cwsusy\ \wrtt extension by $\sw$ (again with 
proper account for stabilizers), but without
restriction on the value of $Q_\sw$. The set of \bc s with fixed \auto\ type 
A$_\theta$ then corresponds to orbits $[\Lambda,\psi]''$, with the 
same minimal model and fermion alignment selection and identification rules 
as for $\theta\eq0$, but with different charge projection condition 
  \be  Q_\sw(\Lambda) = \theta/\pi\bmod\zet \,.  \ee
Hereby in particular the counting of all A$_\theta$-type \bs s is
reduced to simple, if lengthy, combinatorics, which can
(and should) be directly implemented in a computer algorithm.

\sect{Remarks on B-type boundary conditions}

So far we have restricted our attention exclusively to A-type \bc s. Recall that
these need not preserve all of the chiral \alg\ \ACY\ of the \CY extension, 
but at least its sub\alg\ \Awsusy, which in turn contains \Ainner, the chiral 
\alg\ of the unextended inner sector \Cinner\ (see the chain \erf{extent} 
of embeddings). Now we rather want to study \bc s that possess a
non-trivial glueing condition already for the subalgebra \Ainner.
To have a well-defined total supercurrent for the tensor product, and
for compatibility with the fermion alignment, we must use the same \auto\
of the \nn\ \alg\ for each factor of the tensor product \Cinner.
As mentioned in subsection \ref{sica}, generically the automorphism group we
have to consider is the non-connected Lie group O$(2)$.
In the connected component of the identity of O$(2)$ we can only use the 
identity element $\omega\eq\id$ itself; this gives rise to all A-type
\bc s, of all subtypes discussed above. Here we are interested in automorphisms
from the other connected component, which are characterized by
  \be  J_n\mapsto -J_n \,, \qquad  G^+_r\mapsto \eE^{\ii\gamma} G^-_r \,,
  \qquad  G^-_r\mapsto \eE^{-\ii\gamma} G^+_r \ee
with $\gamma\iN{\dl R}$. Unlike in the case of the identity component,
where only the identity could be chosen, we can allow for any arbitrary
value of $\gamma$. Boundary conditions obtained this way
are referred to as B-{\em type\/} conditions. As a distinguished subset, they
include those where the breaking is induced by the {\em mirror \auto\/}
of the total \nn \alg, which is obtained for $\gamma\eq0$:
  \be  J_n\mapsto -J_n \,,\qquad  G^\pm_r\mapsto G^\mp_r \,. \Labl41
We will use the name \Be-{\em type\/} 
for these specific \bc s; the subscript $C$ reminds of charge conjugation.

As compared to A-type conditions, we have to face a new problem, which also 
arises in other circumstances. Namely, we are only given an automorphism 
of the sub\alg\ \Ainner, but not an automorphism of the full chiral \alg\
\ACY\ of our interest. We have already seen above in the case $\omega\eq\id$
that there may exist several different lifts of an automorphism to a
larger \alg. But at a more fundamental level, there is even no guarantee that 
any of the automorphisms of B-type can be lifted at all.
Indeed, as explained in appendix \ref{aA}, in a general simple current extension
there can exist an obstruction to lifting a given \auto.

Fortunately, in the specific situation of interest to us here, there is in
fact no such obstruction. This follows from the fact \cite{oooy} 
that A- and B-type conditions get exchanged by the mirror map.
In short, the B-type conditions of a \gemo\ are well-defined for the full
symmetry \alg\ \ACY\ because they correspond to the A-type conditions of
the mirror model, which by the results above are fully under control.
In the bulk, the mirror map just amounts to applying charge conjugation in 
the inner sector. But at least in many \gemo s it can also be described 
alternatively \cite{grpl} by forming the modular invariant that is associated
\cite{scya6} to a suitable group of simple currents of non-integral conformal 
weight. These currents implement the `phase symmetries' of the \mimo s. (In 
the literature \cite{grpl,aslu1,schi7,GreE}
this is again usually referred to as an orbifold construction.) Accordingly,
the methods of \cite{prss3,fuSc5,huss2}, which show how to deal 
with \bc s for torus \parfu s associated to simple currents of half-integral 
conformal weight, should be helpful for analyzing the B-type conditions.

But still the concrete details in the description of the B-type conditions are 
rather involved. When formulated in terms of the chiral algebra \Ainner, the
complication manifests itself in the fact that the subalgebra that is preserved
by the boundary conditions is neither contained in \Ainner\ nor does it contain
\Ainner. Roughly, one {\em simultaneously\/} tries to {\em reduce\/} the \alg\ 
-- by taking the orbifold with respect to the \auto\ group $\Gamma$ in
question, e.g.\ \wrt the $\zet_2$ generated by the mirror automorphism \Erf41 --
and to {\em extend\/} it -- by the simple current extension with the group 
$\calgcy$. More concretely, we want to know a lift $\hat\Gamma$ of the orbifold
group $\Gamma$ to \ACY\ and a chiral \alg\ extension $\hat{\cal K}$ 
of $(\Ainner)^\Gamma$ to ${(\ACY)}^{\hat\Gamma}$ such that the diagram 
  \be  \begin{array}{c}
  \hsp{.4}\ACY \\{}\\[-1.2em]
  {{\scs\calgcy}}\!\!{\raisebox{-.45em}{\Large$\nearrow$}} \hsp{3.67} 
  {\raisebox{-.45em}{\Large$\searrow$}}\!\!\!{\scs\hat\Gamma}\hsp{1.77} \\{}\\[-.68em]
  \hsp{1.5}\Ainner \ \,{-}\,{-}\,{-}\,{-}\,{-}\,{\to} \ {(\ACY)}^{\hat\Gamma}
  \\{}\\[-.54em]
  {{\scs\Gamma}}\!\!{\raisebox{.25em}{\Large$\searrow$}} \hsp{3.7}
  {\raisebox{.25em}{\Large$\nearrow$}}\!\!\!{{\scs\hat{\cal K}}}\hsp{.3} \\{}\\[-.98em]
  (\Ainner)^\Gamma\hsp{.12}
  \eear \ee
is commutative. Then in particular the map indicated by the dashed arrow is well 
defined.
On the other hand, for our study of \bc s the crucial issue is the relation between
the largest ($\ACY$) and smallest ($(\Ainner)^\Gamma$) chiral \alg\ in the diagram.
The latter is the orbifold of the former by some \auto\ group $\Gamma\cy$, and
conversely, $\ACY$ is a certain extension $\cal E$ of $(\Ainner)^\Gamma$:
  \be  \begin{array}{rc}
  & \ACY \\{}\\[-1.4em]
  {{\scs\calgcy}}\!\!\!{\raisebox{-.45em}{\Large$\nearrow$}}\hsp{-.98}\\{}\\[-.9em]
  \Ainner\hsp{.3}  \\{}\\[-.9em]
  {{\scs\Gamma}}\!\!\!{\raisebox{.25em}{\Large$\searrow$}}\hsp{-.98}  \\{}\\[-1.2em]
  & \ (\Ainner)^\Gamma                    \\{}\\[-6.6em]
  & \hsp{.41}\mbox{\LARGE$ | \uparrow$}   \\{}\\[-1.7em]
  & \mbox{\LARGE$ \downarrow |$}          \\{}\\[-3.5em]
  & {\scs\Gamma\cy} \hsp{3.22}            \\{}\\[-2.1em]
  & \hsp{2.4}{\scs\cal E}                 \\[3.8em] \eear \ee
Note that $\Gamma\cy$ is not the direct product $\Gamma\,{\times}\,\calg$ of 
the orbifold group $\Gamma$ (here $\zet_2$) and the simple current group 
$\calg$ (here $\calgcy$). In fact, the current $\sv$ \Erf sv is not invariant 
under the mirror \auto, so that even if one dealt with a group 
constructed out of $\Gamma$ and $\calg$, it could definitely not be their direct 
product. Closer inspection shows that the mirror orbifold $(\Ainner)^\Gamma$ 
of the tensor product theory is an orbifold of the \CY extension $\ACY$
by a {\em non-abelian\/} group $\Gamma\cy$. Conversely, in order to obtain the
orbifold of \CCY\ from the mirror orbifold of \Cinner\ one must consider an
extension $\cal E$ of the chiral algebra by fields not all of which are simple 
currents, though still all have integral quantum dimension. One may expect that
$\hat\Gamma$ is a subgroup of $\Gamma\cy$, and that the fields in $\hat{\cal K}$ 
form a subset of those in $\cal E$. Some of the necessary mathematical tools for
attacking this problem have recently been established (see 
e.g.\ \cite{brug2,boev3,boek,muge6,sawi2}), but they are not sufficient to 
obtain a complete description of the associated B-type boundary conditions.

However, we can still draw some general conclusions about B-type conditions 
by again invoking mirror symmetry. As a matter of fact, the statement that 
A- and B-type conditions get exchanged by the mirror symmetry
has to be refined, because the mirror \auto\ \Erf41
only involves the total \nn \suco \alg\ of the tensor product,
whereas as noticed above for the complete specification of A-type (and
likewise of B-type) \bc s we have to prescribe the action of the \auto\
on the full chiral \alg\ of the \CY extension.\,%
 \futnote{In addition, one should be aware of the fact that initially the mirror
 map refers to the situation {\em before\/} applying the bosonic string map.
 However, the mirror map is compatible with the bosonic string map, so that
 we can directly apply it here. The point is that,
 in \cft\ terminology, the mirror map is nothing but charge conjugation.
 But as a consequence of formula \Erf st we have $\tilde C B \eq B C$,
 where $C$ is the charge conjugation matrix of the bosonic theory (i.e., after
 the bosonic string map $B$) and $\tilde C$ the charge conjugation matrix of
 the supersymmetric theory (before the bosonic string map).}
To do so, we use the fact that, as a special case of more general T-duality 
relations \cite{fuSc12}, those \bc s of a \cft\ with torus \parfu\ $Z$ which 
leave invariant precisely the subalgebra of the chiral \alg\ that is pointwise 
fixed under the charge conjugation \auto\ $\omega^C$ can be put in one-to-one 
correspondence with the \bc s that preserve the full chiral \alg\ of a 
different \cft, namely the one with torus \parfu\ $CZ$.\,%
 \futnote{The simplest instance of this phenomenon is the exchange between
 Neumann and Dirichlet conditions in the theory of a single free boson.}
It follows that the \Ae-type \bc s of a \gemo\ correspond to the \Be-type 
\bc s of its mirror model, and vice versa. How this generalizes to the more 
general A- and B-type conditions will be discussed elsewhere. In particular,
this consideration tells us that there is no obstruction to a lift of the mirror
\auto\ of the total \nn \alg\ to the full chiral \alg\ of the \CY extension.

Finally let us mention that whenever the tensor product \Cinner\ contains 
identical factors, there are additional \auto s, beyond those belonging to O(2) 
discussed above, that can be modded out without spoiling world sheet \susy, 
namely the permutation symmetries that interchange the identical factors.

\sect{Fixed points and singularities}

We conclude this paper with a few comments on the relationship between fixed 
point structures in Gepner models and various other `singular' structures 
that occur in the analysis of string compactifications on \CY spaces and in 
Gepner models. The following list summarizes various such structures.
\nxt
It is generally believed that a Gepner model describes the exact 
solution of string theory compactified on a certain \CY manifold, at a 
specific point of its moduli space. There is a general prescription for 
finding the polynomial constraints that provide the embedding of the \CY
manifold in a weighted projective space, see e.g.\ \cite{grvw,fkss2,witt41}.
When carrying this construction through, one encounters the problem that 
the variety defined by those polynomial constraints is not smooth, but has
singularities. It is only after resolving these singularities that one obtains
the \CY manifold.
\nxt
The moduli of the \CY manifold are related to the $(c,c)$ and $(a,c)$ rings
of the \nn superconformal field theory \cite{levw}. For instance, 
deformations of the complex structure of the \CY space correspond to
Gepner model fields that are chiral primaries with respect to both the
left- and right-moving chiral algebras, with total $\uone$ charges $1$. Fields 
that have in addition identical left- and right-moving labels in each \mimo\ 
can be related to polynomial deformations of the complex structure, i.e.\ to 
a change in the defining polynomial. But it has been pointed out long ago
\cite{grhu2} that polynomial deformations give neither a complete nor
an unambiguous enumeration of complex structure deformations.
In the Gepner model, this can be related to the existence of twisted, i.e.\
left-right asymmetric, modes in the $(c,c)$ ring, see e.g.\ \cite{aslu}.\,%
 \futnote{These states can be obtained directly by the algorithms incorporating
 the simple current extension, e.g.\ by the computer programs that were
 used to produce the bulk spectra in \cite{scya4,fkss2,aslu,sche3,fuSc}.}
\nxt
In the study of boundary states in Gepner models
and their comparison with geometric D-branes and bundles in the corresponding
\CY manifolds \cite{bDlr,diRo,kllw,scHe',nano}, it was noticed
that several boundary states constructed in \cite{reSC} are not elementary,
in the sense that the annulus coefficient of the vacuum field is larger than
one \cite{diRo}. In \cite{kllw} it has been argued that this can be
understood if one assumes that the relevant
boundary states correspond to branes carrying reducible bundles.

There have been speculations that (some of) these different singular 
structures are intimately related. Now in order for any
definite relationship between geometrical and \cft\ structures to
exist, at least the combinatorial data characterizing them should be
similar. In the situation at hand, this is actually not the case.
\nxt
Let us first recall some of our results concerning A-type \bc s.  
As we have explained, the construction of
A-type \bs s in Gepner models is completely under control, and only
existing technology \cite{card9,fusS6,fuSc11,fuSc12} is needed.\,%
 \futnote{In particular, the fixed points can already be understood entirely
 at the chiral level in closed string theory. It is therefore e.g.\ 
 unnecessary to study open string partition functions in order to fix
 normalization factors (as advocated in \cite{reSC,brSc}).
 Rather, the annulus coefficients are non-negative integers by construction.
 Also, nothing is gained by separating boundary states or partition functions
 into a part from an `untwisted' and one from a `twisted' sector; all the
 relevant structures are already present in the closed string theory.}
We have also seen that the fixed points in Gepner models are always of $\zet_2$ 
type, i.e.\ the only non-trivial stabilizer group that occurs is of the form 
$\{\vac,\L\}$. On the other hand, singularities in the construction of the 
\CY manifold as a complete intersection in a weighted projective space can 
occur if the weights have common divisors, and are locally of the type
$\complex^d/\zet_\ell$ where $d\eq2,3$ and $\ell$ can be any
integer, see, for example, \cite{hkty}.
\nxt
It is also unlikely that there is any relationship
between twisted modes in the $(c,c)$ ring and singularities in the
construction of the \CY space. By inspecting lists of twisted
$(c,c)$ fields for particular Gepner models, it is easily checked
that they do not display any $\zet_\ell$ structure. And indeed, the
existence of non-polynomial deformations is not related to a
singularity occuring in the construction of the manifold, but to
the presence of an obstruction in the relevant cohomologies of the family
of smooth manifolds \cite{grhu2}.  On the \cft\ side, twisted modes may
also be related to the presence of an enhanced symmetry. For instance,
compactification on K3 leads to $N\eq4$ world sheet \susy, and half of the 
modes in the $(c,c)$ and $(c,a)$ rings are twisted. Incidentally, in such a 
situation boundary conditions might break different parts of this extended 
symmetry, so that a finer classification of boundary conditions is possible.
\nxt
Furthermore, fixed points in the Gepner extension do not
seem to be related to the presence of twisted modes in the $(c,c)$ 
ring. In fact, the Gepner extension has fixed points whenever at least
one level is even. But many of those Gepner models do not possess
any twisted modes in their $(c,c)$ ring. A simple example is provided
by the Gepner model $(2,2)$ which does have fixed points, but simply 
corresponds to a torus embedded
in one-\dim al projective space, without any singularities.

Finally we would like to point out that in our opinion several interesting
problems concerning boundary states in Gepner
models and branes on \CY compactifications are still open.
For example, the construction of states charged under
twisted modes in the $(c,c)$ ring, as well as a better
understanding of composite boundary states and their connection
to reducible bundles, is highly desirable.

\appendix
\section{Lifting orbifold \auto s} \label{aA}

In the main text we have seen that when analyzing B-type \bc s one must
study the interplay between simple current extensions and orbifolds. 
Here we investigate this issue, which is of interest also in other
situations, in its own right.
Thus consider an arbitrary \rcft\ with chiral \alg\ $\cala$, a group $\calg$
of simple currents (with integral conformal weight) of the theory, and
a group $\Gamma$ of \auto s of $\cala$. Every \auto\ $\omega$ of $\cala$
induces a permutation of the (labels for) primary fields, which 
is an \auto\ of the fusion rules and which we denote by $\omega^*$.
By including the simple currents into the chiral \alg, one obtains an
extended theory with chiral \alg\ $\cala\ext\,{\supset}\,\cala$, while
by dividing out the \auto s in $\Gamma$ one obtains an orbifold
theory with chiral \alg\ $\cala^\Gamma\,{\subset}\,\cala$.
Our goal is then to make sense of the symbol $\cala^\Gamma\ext$. For 
simplicity and definiteness we restrict to the case $\Gamma\eq\zet_2$, i.e.\ 
there is only a single non-trivial \auto\ $\omega$ and it has order two.

A necessary prerequisite for attaining this goal is to lift the \auto\ 
$\omega$ to some \auto\ $\omega\ext$ of the extended chiral \alg\
$\cala\ext$. As a matter of fact, our first task should be to determine 
whether such a lift is possible at all. Indeed there can be an obstruction, 
and this will be studied below. But for the moment let us restrict 
to those cases where the lifting of $\omega$ is not obstructed. 
In that case we need to investigate the uniqueness of the lift.
We already know from the discussion in the main text that typically
even the identity \auto\ of $\cala$ will possess
several distinct lifts to $\cala\ext$. To be more concrete,
we use the fact that we can characterize \cite{fuSc11}
$\cala$ as the fixed algebra of $\cala\ext$ under an action of the
group $G\ext\,{:=}\,\calg^*$ of characters of the simple current group
$\calg$. Further, $\cala^\omega$ is the fixed algebra in $\cala$
under $\omega$. Thus we can characterize $\cala^\omega$ equally well
as the subalgebra of the big algebra $\cala\ext$ under the combined action
of $G\ext$ and $\omega\ext$, i.e.\ under the action of the group $H\ext$
generated by $G\ext$ and $\omega\ext$.
Note that we do {\em not\/} assume here that $\omega\ext$ has order two.

We can then again employ the result from the Galois theory for vertex operator
algebras that the possible chiral algebras between $\cala^\omega$ and 
$\cala\ext$ are in one-to-one correspondence with subgroups of the group $H\ext$. 
Now since $\cala$ is the fixed point subalgebra of $\cala\ext$ under $G\ext$,
for every $g\iN G\ext$ the element 
$g_\omega\,{:=}\,\omega\ext^{-1}{\cdot}g{\cdot}\omega\ext\iN H\ext$ 
acts on $\cala$ as $\omega^{-1}{\cdot}1{\cdot}\omega\eq1$, and
by the Galois correspondence every such $g_\omega$ already lies in $G\ext$.
This tells us that $G\ext$ is a normal abelian subgroup of $H\ext$,
and the conjugation by $\omega\ext$ acts on $G\ext$ by an outer
automorphism $\hat\omega$, i.e.\ $\omega\ext^{-1}\, g\, \omega\ext\eq
\hat\omega(g)$. As $\omega\ext^2\iN G\ext$ and $G\ext$ is abelian,
the automorphism $\hat\omega$ has order two. So we have the structure\,%
 \futnote{It is instructive to think of $H\ext$ like of a compact Lie
 group with two connected components. The `identity component' is
 $G\ext$, and it has a natural unit element, while the other component,
 consisting of the automorphisms of $\cala\ext$ whose restriction to
 $\cala$ acts like $\omega$, does not possess a natural base point.}
  \be  0 \,\to\, G\ext \,\to\, H\ext \,\to\, \zet_2 \,\to\, 0 \,.  \ee
As $\omega\ext$ does not necessarily have order two,
the term `orbifold of $\cala\ext$ by $\omega\ext$' is to be interpreted
as the orbifold by the cyclic subgroup of $H\ext$ that is generated by
$\omega\ext$. But those elements of this subgroup that are of the form
$(\omega\ext)^{2n}$ form a cyclic subgroup $G\ext^0$ of $G\ext$,
so that we may as well perform the orbifolding stepwise, first by
$G\ext^0$ and afterwards by $\omega\ext$ which then has order two on
$(\cala\ext)^{G\ext^0}$. It follows that at the price of possibly working 
with a different simple current extension than the original one, we may 
restrict our attention to the case when $\omega\ext$ has order two.

A simple illustration of the non-uniqueness of the lift is provided by
the following $c\eq1$ theories. The original theory is the rational free
boson $X$ at compactification radius $R^2\eq mn^2$ with $n$ integral and $m$
even integral. Thus the chiral \alg\ $\cala$ is generated by the current
$j\eq\ii\partial X$, giving rise to $\uone_{mn^2}$, and by the two fields
$\Phi_\pm\eq\exp(\pm\ii\sqrt m nX)$; there are $mn^2$
primary fields, which we label by the integers from 0 to $mn^2{-}1$
and each of which is a simple current. Extending this theory by the simple
currents $\calg\eq\{\ell mn\,|\, \ell\eq0,1,...\,,n{-}1\}\,{\cong}\,\zet_n$,
one obtains the theory of a free boson with compactification radius $R^2\eq m$
and chiral \alg\ $\cala\ext$ generated by $\uone_m$ and $\Phi^{\rm ext}_\pm\eq
\exp(\pm\ii\sqrt m X)$. On the other hand, dividing out the charge
conjugation \auto\ $\omega$ from $\cala$ one arrives at the $\zet_2$ orbifold
of the free boson, with $mn^2/2\,{+}7$ primary fields.
The map $\omega$ acts on the fields generating $\cala$ as
  \be  \omega(j) = -j \,, \qquad  \omega(\Phi_\pm) = \Phi_\mp \,.  \Labl9o
It can be lifted to $\cala\ext$ as
  \be  \omega\ext(j) = -j \,, \qquad  \omega\ext(\Phi^{\rm ext}_\pm)
  = \zeta^{\pm1}\, \Phi^{\rm ext}_\mp \,,  \Labl+2
where $\zeta$ is an arbitrary $n$th root of unity. In terms of the free boson
$X$, this reads
  \be  \omega\ext(X) = -X + \Frac{2\pi\ell}{n\sqrt m}  \ee
with some $\ell\iN\{0,1,...\,,n{-}1\}$.
This example also displays nicely that even the identity map of $\cala$ can
typically be lifted in several inequivalent ways. Clearly, for each
$n$th root $\zeta$ of unity, the map
  \be  \id\ext(j) = j \,, \qquad  \id\ext(\Phi^{\rm ext}_\pm)
  = \zeta^{\pm1}\, \Phi^{\rm ext}_\pm   \Labl-2
(acting on the free boson field as $X\Mapsto X\,{+}\,2\pi\ell/n\sqrt m$ for
$\zeta\eq\exp(2\pi\ii\ell/n)$) is an \auto\ of $\cala\ext$ and restricts to
the identity map on $\cala$. Note that while the order of the map given by
\Erf+2 is 2 independently of the value of $\zeta$, this is no longer true
for the map \Erf-2.

Let us now come back to the issue of existence of $\omega\ext$. Under the 
most general circumstances such a lift may actually not exist at all.  
First, clearly the compatibility condition
$\omega^*(\calg)\eq\calg$ must be satisfied. In the sequel we assume that
this is the case. (In the situation of our interest, this condition is indeed
met. Also, the condition is automatically fulfilled whenever an extension is
by {\em all\/} integer spin simple currents of a given theory.) But even with 
this assumption the existence of a lift $\omega\ext$ is not guaranteed.
Rather, one has to study the relation between the subgroup
  \be \calg_0 := \{ \J\iN\calg \,|\, \omega^*(\J)\eq\J \} \Labl c0
of $\calg$ and the group $\calg^\omega$ of simple currents of the 
$\cala^\omega$-theory.  By general orbifold rules \cite{dvvv,bifs},
each $\J\iN\calg_0$ gives rise to two simple currents
$\J_\pm$ in the untwisted sector of the orbifold theory $\cala^\omega$.
The fields $\J_\pm$ form a subgroup $\calg^\omega_0$ of $\calg^\omega$.
Thus $\calg^\omega_0$ is a $\zet_2$-extension of $\calg_0$, i.e.\
there is an exact sequence
  \be  0 \,\to\, \zet_2 \,\to\, \calg_0^\omega \,\stackrel\pi\to\,
  \calg_0 \to 0 \,, \labl{exseq}
where the projection $\pi$ acts as $\pi(\J_+)\eq\J\eq\pi(\J_-)$.
But $\calg^\omega_0$
is not necessarily a direct product of $\calg_0$ with $\zet_2$;
the obstruction is expressed by an element $[\eps]$ in $H^2(\calg_0,\zet_2)$.

Inspecting the fusion rules among the fields $\J_\pm$ in the
orbifold theory $\cala^\omega$, one finds that this cohomology class 
$[\eps]$ has the following \cft\ interpretation. The associated 
{\em commutator cocycle\/} $\bar\eps(\J_1,\J_2)\,{:=}\,\eps(\J_1,\J_2)/
\eps(\J_2,\J_1)$ (which only depends on the class $[\eps]$ and not on 
the choice of a representative $\eps$) can be expressed as
  \be  \bar\eps(\J_1,\J_2) = \sum_{\dot\mu} 
  S^{(0)}_{\J_1,\dot\mu} S^{(0)}_{\J_2,\dot\mu}
  (S^{(0)}_{\J_1\J_2,\dot\mu})^*_{} \,/\, S^{(0)}_{\Omega,\dot\mu} \,.  \ee
Here $S^{(0)}$ denotes the matrix that governs the modular behavior of the 
differences $\chii_{\mu_+}{-}\chii_{\mu_-}$ of orbifold characters in the 
untwisted sector coming from the same $\cala$-primary. More specifically, 
under $\tau\Mapsto{-}1/\tau$ these differences become linear combinations
of characters $\chii_{\dot\mu}$ in the twisted sector, with the coefficients
given by $S^{(0)}$ (for more details, see \cite{bifs}).
By the consistency of the orbifold fusion rules, $\bar\eps$ can only take the
values $\pm1$ and satisfies $\bar\eps(\J_1,\J_2)\eq\bar\eps(\J_2,\J_1)$,
and therefore \cite{KArp4} indeed determines a unique class $[\eps]$ in 
$H^2(\calg_0,\zet_2)$. In the special case where $\calg_0\eq\calg$,
for the construction of $\cala^\omega\ext$ we must pick, for each $\J\iN\calg$,
one of the fields $\J_\pm\iN\calg^\omega$ of the $\omega$-orbifold, in such
a manner that the chosen set of representatives closes under fusion. Thus
we must find a section $\sigma{:}\; \calg^0\,{\to}\,\calg^\omega$ for the exact 
sequence \erf{exseq}. Such a section exists only if the extension is trivial, 
i.e.\ if $[\eps]\eq1$.  In conclusion, there is an obstruction to the lift 
of $\omega$ to an automorphism $\omega\ext$ of $\cala\ext$; moreover,
it is controlled by the twisted sector of the orbifold, and hence computable.

For various classes of orbifold constructions the presence of an
obstruction can be decided without too much effort.
For instance, from the results of \cite{bifs} it follows immediately that
there is no obstruction when $\omega$ is an \auto\ of the chiral \alg\
of a \wzwm\ that comes from an inner \auto\ of the underlying simple \lie.
Similarly, no obstruction is present
for the charge conjugation orbifold of a single free boson\,%
 \futnote{When $n$ as introduced before formula \Erf9o
 is odd, then $\calg_0\eq\{0\}$, and the statement is trivial.
 When $n$ is even, then $\calg_0\eq\{0,mn^2/2\}\,{\cong}\,\zet_2$, while the
 simple current group $\calg^\omega$ of the orbifold is $\zet_2\,{\times}\,
 \zet_2$, so $\calg^\omega$ is a trivial extension of $\calg_0$.}
and for arbitrary permutation orbifolds. As a matter of fact,
we do not know of any orbifold construction where the obstruction is present.
It is tempting to expect that the obstruction is indeed absent in all cases
that appear in \cft, but so far we do not have any general argument to this
effect. In any case, in all the applications in the main text we are able to 
show that the obstruction is absent.
Therefore in the present paper we do not attempt to push this issue further.

The fields in $\calg$ which are not contained in the subgroup $\calg_0$
\Erf c0 come in pairs $\J$ and $\omega^*(\J)$, and each such pair
gives rise to a single field in $\cala^\omega$, which has quantum dimension
2, i.e.\ is {\em not\/} a simple current any longer. As a consequence, these
fields do not have a direct influence on the presence of an obstruction.  
On the other hand, even when there is no obstruction, it turns out to be
quite non-trivial to describe such non-simple current fields in sufficiently
explicit terms in concrete models. In particular, consistency of the 
fusion rules of the $\cala$-theory does not seem to be of any help.

\bigskip\bigskip\noindent{\small {\sc Acknowledgement} \\
We are grateful to Jens Fjelstad, J\"urg Fr\"ohlich, Lennaert Huiszoon,
Peter Kaste, Wolfgang Lerche, Andy L\"utken, and Bert Schellekens for 
helpful discussions and comments.}

\newpage
\small
 \newcommand\wb{\,\linebreak[0]} \def\wB {$\,$\wb}
 \newcommand\Bi[1]    {\bibitem{#1}}
 \renewcommand\J[5]   {{\sl #5}, {#1} {#2} ({#3}) {#4} }
 \newcommand\PhD[2]   {{\sl #2}, Ph.D.\ thesis (#1)}
 \newcommand\Prep[2]  {{\sl #2}, preprint {#1}}
 \newcommand\BOOK[4]  {{\em #1\/} ({#2}, {#3} {#4})}
 \newcommand\inBO[7]  {in:\ {\em #1}, {#2}\ ({#3}, {#4} {#5}), p.\ {#6}}
 \def\jf    {J.\ Fuchs}
 \def\adma  {Adv.\wb Math.}
 \def\anop  {Ann.\wb Phys.}
 \def\aspm  {Adv.\wb Stu\-dies\wB in\wB Pure\wB Math.}
 \def\atmp  {Adv.\wb Theor.\wb Math.\wb Phys.}
 \def\comp  {Com\-mun.\wb Math.\wb Phys.}
 \def\duke  {Duke\wB Math.\wb J.}
 \def\ijmp  {Int.\wb J.\wb Mod.\wb Phys.\ A}
 \def\joal  {J.\wB Al\-ge\-bra}
 \def\maan  {Math.\wb Annal.}
 \def\mpla  {Mod.\wb Phys.\wb Lett.\ A}
 \def\nuci  {Nuovo\wB Cim.}
 \def\nupb  {Nucl.\wb Phys.\ B}
 \def\phlb  {Phys.\wb Lett.\ B}
 \def\phrl  {Phys.\wb Rev.\wb Lett.}
 \def\phrp  {Phys.\wb Rep.}
 \def\NH     {{North Holland Publishing Company}}
 \def\SV     {{Sprin\-ger Ver\-lag}}
 \def\WS     {{World Scientific}}
 \def\Ad     {{Amsterdam}}
 \def\Be     {{Berlin}}
 \def\Si     {{Singapore}}

\end{document}